\documentclass[twocolumn,pre,twoside,groupedaddress,floatfix]{revtex4}

\usepackage{amsmath}
\usepackage{amssymb}
\usepackage{graphicx}
\usepackage{dcolumn}
\usepackage{bm}
\usepackage{color}

\def\Re{\mathop\mathrm{Re}}

\begin{document}

%-------------------------------------------------------------------------------

\title{Impedance spectroscopy of ions at liquid-liquid interfaces}

\author{Andreas Reindl}
\email{reindl@is.mpg.de}
\author{Markus Bier}
\email{bier@is.mpg.de}
\affiliation
{
   Max-Planck-Institut f\"ur Intelligente Systeme, 
   Heisenbergstr.\ 3,
   70569 Stuttgart,
   Germany, 
   and
   Institut f\"ur Theoretische Physik IV,
   Universit\"at Stuttgart,
   Pfaffenwaldring 57,
   70569 Stuttgart,
   Germany
}

\date{01 October, 2013}

\begin{abstract}
The possibility to extract properties of an interface between two immiscible 
liquids, e.g., electrolyte solutions or polyelectrolyte multilayers,
by means of impedance spectroscopy is investigated theoretically within a 
dynamic density functional theory which is equivalent to the Nernst-Planck-Poisson 
theory.
A novel approach based on a two-step fitting procedure of an equivalent circuit
to impedance spectra is proposed which allows to uniquely separate bulk
and interfacial elements.
Moreover, the proposed method avoids overfitting of the bulk 
properties of the two liquids in contact and underfitting of the interfacial
properties, as they might occur for standard one-step procedures.
The key idea is to determine the bulk elements of the equivalent circuit in a
first step by fitting corresponding sub-circuits to the spectra of uniform 
electrolyte solutions, and afterwards fitting the full equivalent circuit
with fixed bulk elements to the impedance spectrum containing the interface.
This approach is exemplified for an equivalent circuit which leads to a physically
intuitive qualitative behavior as well as to quantitively realistic values of the 
interfacial elements.
The proposed method is robust such that it can be expected to be applicable to
a wide class of systems with liquid-liquid interfaces.
\end{abstract}

\maketitle

%-------------------------------------------------------------------------------

\section{\label{sec:intro}Introduction}

Impedance spectroscopy, which is concerned with the frequency dependence of the
linear current-voltage response, is a well-known standard technique to 
investigate electrode processes, e.g., electrochemical reactions
\cite{Macdonald1992,Bagotsky2006,Schmickler2010}.
Moreover, impedance spectroscopy has been used to experimentally study
interfaces between two immiscible electrolyte solutions (ITIES) 
\cite{Samec1981, Hajkova1983, Geblewicz1984, Samec1987, Wandlowski1988,
Samec2004, Vanysek2008} as well as between an electrolyte solution
and a polyelectrolyte multilayer (PEM) \cite{Barreira2004, Silva2005, Ruan2009,
Patricio2010}.
It is common to analyze measured impedance data by fitting equivalent circuits 
composed of, e.g., Ohmic resistors and capacitors, and identifying the
corresponding elements by microscopic processes.
However, equivalent circuits are not uniquely determined by their impedance
spectrum \cite{Macdonald1992}, i.e., a particular
choice has to be motivated \cite{Franceschetti1977}, e.g., by physical arguments.

In the cited studies of ITIES the impedance spectra are fitted by
an Ohmic resistor representing the bulk phases in series with a Randles-type 
equivalent circuit \cite{Randles1947} corresponding to the interfacial properties.
This separation is justified by the observation that the bulk relaxation times
of molecular solvents are several orders of magnitude shorter than those of the 
interfacial processes.
Since in these studies the bulk phases were many orders of magnitude larger
than the Debye lengths the Ohmic bulk resistance could be determined rather
precisely \cite{Macdonald1977}.

The situation is different for the cited studies of interfaces
between molecular solvents and PEMs, where both phases can
be considerably larger than the Debye lengths, but the relaxation times of the
PEMs can be substantially longer than for molecular solvents \cite{Vyas2010}
so that PEM bulk processes occur at frequencies which are not separate from
frequencies of interfacial processes.
Consequently, the equivalent circuits used in these studies did not attempt 
to account for the PEMs just by an Ohmic resistor alone but also by some 
capacitive behavior.
However, the equivalent circuits used there to describe the PEM bulk and the
interfacial processes are of the type initially designed for systems without
interfaces \cite{Franceschetti1977} so that PEM bulk and interfacial processes
are not well separated in these analyses, which renders the interpretation of
the equivalent circuit elements difficult.
Another, more technical, aspect is that fitting equivalent circuits containing
elements representing both bulk and interfacial processes to measured impedance
spectra in one step poses the risk of overfitting the bulk elements at the 
expense of underfitting the interfacial elements.
This may result in inaccurate values of the interfacial
elements. In the past a similar observation on systems without a liquid-liquid interface
has led to the advice to ``use a priori information when available'' \cite{Macdonald1977}.

In the present publication we propose a novel, alternative approach to extract
interfacial properties from impedance spectra, which does not exhibit
the problems occurring in the studies of the electrolyte solution / PEM 
interfaces cited above.
It consists of two steps in the first of which the subcircuits of the full
equivalent circuit which describe the bulk phases in contact are 
fitted to impedance spectra of the corresponding \emph{uniform} systems without
liquid-liquid interfaces.
In the second step the interfacial elements are determined by fitting the full
equivalent circuit where the previously determined parameters of the bulk 
elements are fixed.
This two-step procedure is in analogy to the determination of an interfacial 
tension by subtracting the bulk contribution from the total free energy.
The proposed two-step method may be also viewed as an analog to the
approach determining a signal on top of a background by first measuring the
background without signal and afterwards subtracting the background from the 
data containing the signal.
The proposed two-step fitting procedure solves the problem of bulk and interfacial
processes being lumped in the same equivalent circuit elements since in the second
step an element is either fixed (bulk) or free to be fitted (interfacial).
Moreover, by separating fitting bulk and interfacial elements overfitting of the
former and underfitting of the latter does not occur.
In order to demonstrate this method and to assess the quality of the resulting
interfacial parameters, we introduce in Sec.~\ref{sec:Formalism} a simple
model of an electrolytic cell which is transparently analyzed within a dynamic
density functional approach that is equivalent to the Nernst-Planck-Poisson 
theory.
We thereby imply solvents which lead to conventional diffusion of ions, 
whereas solvents giving rise to, e.g., anomalous diffusion \cite{Macdonald2010} are
not considered here.
After introducing a possible equivalent circuit for bulk liquids the 
interfacial parameters, which have been determined according to the two-step 
method described above, are discussed in Sec.~\ref{sec:Discussion}.
Finally conclusions are given in Sec.~\ref{sec:Conclusion}.

\vfill

%-------------------------------------------------------------------------------

\section{\label{sec:Formalism}Formalism}

\subsection{Setup\label{subsec:Setup}}

\begin{figure}[!t]
  \includegraphics[width=0.4\textwidth]{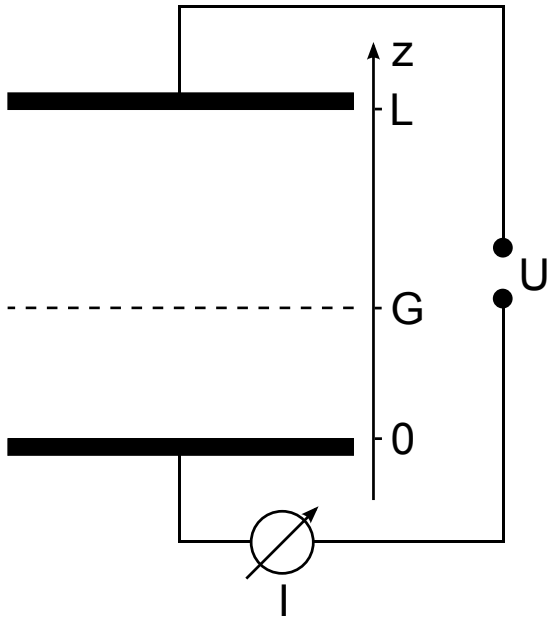}
  \caption{Sketch of the two-electrode setup considered in the present study. 
           A voltage $U$ is applied between the electrodes at positions $z=0$ and $z=L$, 
           which bound an electrochemical cell. The cell may contain up to two immiscible liquid solvents 
           which lead to the formation of a liquid-liquid interface that is drawn at position $z=G$. 
           The electric current $I$ is measured in reply to the applied voltage $U$.}
  \label{fig:sketch}
\end{figure}
Figure~\ref{fig:sketch} displays a sketch of the setup considered here.
An alternating voltage $U(t)$ with frequency $\omega$ is applied to two planar electrodes at
positions $z=0$ and $z=L$, which bound an electrochemical cell. 
For simplicity only a two-electrode system with blocking electrodes is considered. 
However, the method of determining interfacial properties from impedance spectra to be described 
below is generally applicable, e.g., also for four-electrode systems or non-blocking electrodes.
The cell contains either a single solvent or two immiscible solvents, the interface between which is
represented by the dashed line in Fig.~\ref{fig:sketch} located at position $z=G$.
The ions, belonging to two or more ion species, are partitioned amongst the solvents according to
their solubility without being adsorbed or involved in chemical reactions at the electrodes or the
liquid-liquid interface.
Further we assume that the solvents do not flow, neither due to the action of the electric field
onto the solvent dipoles nor due to electro-osmosis. 
The former effect may be avoided by, e.g., a polyelectrolyte multilayer or ion-conducting solids as solvents. 
The latter effect can be expected to be weak anyway because ions of both charge signs are mobile,
i.e., their movement in the electric field leads to a negligible net flow. 
Therefore the dynamics of the system is described by the diffusion of ions in quiescent solvents. 
Given the current $I(t)$ in response to the voltage $U(t)$, the amplitude ratio and the phase 
difference between voltage $U(t)$ and current $I(t)$ determines the (complex-valued) impedance 
$Z(\omega)$.

\subsection{Model}
The impedance $Z$ and the relative permittivity $\eta$ are experimentally accessible quantities.
However, in order to clearly demonstrate our approach of extracting interfacial properties from 
impedance spectra, instead of real experimental data, we discuss impedance spectra calculated 
within a theoretical model. 
Since we consider the case of a purely diffusive conserved ion dynamics (model B) in quiescent
solvents (see Subsec.~\ref{subsec:Setup}), dynamic density functional theory as
formulated in Ref.~\cite{Dieterich1990} is an appropriate theoretical framework.
The Helmholtz free energy in units of $k_BT$ as a functional of the ion densities 
$\varrho_\alpha(z)/l_B^3$, where $\alpha$ denotes the ion species, is given by
\begin{align}
  \begin{aligned}
    \frac{F[\{\varrho_\alpha(z)\}]}{A} = \sum_\alpha \int \limits _{0^+}^{L^-}\text{dz}\, \varrho_\alpha(z)(\ln(\varrho_\alpha(z))-1+V_\alpha(z))\\
                                        + \frac{1}{8\pi}\int \limits _{0^+}^{L^-}\text{dz}\,\epsilon(z)\phi'(z)^2 - (\phi_0\sigma_0 + \phi_L 	\sigma_L),
  \end{aligned}
 \label{eq:free_energy_functional}
\end{align}
where $Al^2_B$ is the cross-sectioned area, $V_\alpha(z)k_BT $ is an arbitrary external potential, 
$\epsilon(z)$ is the relative permittivity of the solvents with
respect to the vacuum permittivity $\epsilon_0$, $\phi(z)k_BT/e$ is the electrostatic potential and $\sigma e/l^2_B$ is the 
surface charge density of the electrodes. 
The indices $\{0,L\}$ denote that the value of the quantity in question is to be taken at 
$z=\{0,L\}$. Lengths are scaled by the vacuum Bjerrum length $l_B=e^2/(4\pi\epsilon_0 k_BT)$.
The first term on the right-hand side of Eq.~(\ref{eq:free_energy_functional}) is the ideal gas
term. 
The electrostatic interaction is accounted for by the second term. The third one takes into account
the free energy contribution of the voltage source.

The electrostatic potential $\phi$ is a functional of the ion densities. 
This relation is specified by Poisson's equation:
\begin{align}
  \begin{aligned}
    -(\epsilon(z)\phi'(z))'=4\pi\sum\limits_\alpha q_\alpha \varrho_\alpha(z)\\
    \text{where}\quad\phi(0)=\phi_0,\quad\phi(L)=\phi_L.
  \end{aligned}
  \label{eq:Poisson_equation}
\end{align}
The $'$ in Eq.~(\ref{eq:Poisson_equation}) denotes derivation with respect to $z$. 
As charges are scaled by the elementary charge $e$, $q_\alpha$ is the valency of each ion species
$\alpha$.

Functional derivation of Eq.~(\ref{eq:free_energy_functional}) with respect to the ion density
gives the chemical potential $\mu_\alpha$:
\begin{align}
  \begin{aligned}
    \mu_\alpha(z,t)&=\frac{\delta F}{\delta\varrho_\alpha(z)}\bigg|_{\{\varrho_{\alpha'}(\cdot,t)\}}\\
                &=\ln(\varrho_\alpha(z,t))+V_\alpha(z,t)+q_\alpha\phi(z,t).
  \end{aligned}
  \label{eq:chemical_potential}
\end{align}
Fick's law 
\begin{align}
  j_\alpha(z,t)=-\Gamma_\alpha(z)\varrho_\alpha(z,t)\frac{\partial\mu_\alpha}{\partial z}(z,t)
  \label{eq:Ficks_law}
\end{align}
relates the negative gradient of the chemical potential, which may be interpreted as a force, to
the current density $j_\alpha/(l^2_B\tau)$, where $\Gamma_\alpha(z)/\Gamma_{ref}$ denotes the
mobility of ions of species $\alpha$ at position $z$. Here $\Gamma_{ref}$ is some reference
mobility, which implies the time scale $\tau=l^2_{B}/(\Gamma_{ref}k_BT)$.
We assume blocking electrodes so the current densities have to be zero at $z=\{0,L\}$:
\begin{align}
  j_\alpha(0,t)=j_\alpha(L,t)=0.
\end{align}
The continuity equation ensures that the amount of ions is locally conserved:
\begin{align}
 \frac{\partial\varrho_\alpha}{\partial t}(z,t)=-\frac{\partial j_\alpha}{\partial z}(z,t).
 \label{eq:continuity_equation}
\end{align}
Equations (\ref{eq:chemical_potential}-\ref{eq:continuity_equation}) may be combined to the
Nernst-Planck equation
\begin{align}
  \frac{\partial\varrho_\alpha}{\partial t}(z,t)=\frac{\partial}{\partial z}\left\{\Gamma_\alpha\left[\frac{\partial\varrho_\alpha}{\partial z}+\varrho_\alpha\left(\frac{\partial V_\alpha}{\partial z}+q_\alpha\frac{\partial\phi}{\partial z}\right)\right]\right\}.
  \label{eq:Nernst_Planck_equation}
\end{align}
Equations (\ref{eq:Poisson_equation}) and (\ref{eq:Nernst_Planck_equation}) are coupled, non-linear
differential equations the solutions of which consist of $\alpha$ ion density profiles
$\varrho_\alpha(z,t)$ and the electrostatic potential $\phi(z,t)$.

\subsection{Spectra}
Once the solutions of Eqs. (\ref{eq:Poisson_equation}) and (\ref{eq:Nernst_Planck_equation}) have 
been determined, one can calculate quantities which allow for comparison with measurements.
To that end we assume harmonically alternating electrode potentials
\begin{align}
  \begin{aligned}
  \phi_0(t) &= -\phi_L(t)=\Phi^{st}+\Phi\cos(\omega t)\\
            &= \Phi^{st}+\frac{1}{2}\big(\Phi\exp(\text{i}\omega t)+cc\big)
  \end{aligned}
  \label{eq:alternating_potential}
\end{align}
of frequency $\omega/\tau$ and with the Fourier coefficient $\hat{\phi}_0(\omega)=\Phi/2$, which may be
shifted by a static (superscript $^{st}$) offset $\Phi^{st}$. 
In Eq.~(\ref{eq:alternating_potential}) ``$+cc$'' means, that the complex conjugate of the previous
expression has to be added. 
The voltage $U(t)$ between the electrodes leads to
\begin{align}
  U(t)&=\phi_0(t)-\phi_L(t)=2\Phi^{st}+\hat{U}(\omega)\exp(\text{i}\omega t)+cc
  \label{eq:alternating_voltage}
\end{align}
with $\hat{U}(\omega)=\Phi$.
Using Gauss' law the surface charge density $\sigma_0(t)$ can be expressed with help of the
electric displacement $D(z,t)$ at $z=0$:
\begin{align}
  \begin{aligned}
    \sigma_0(t) &= -\frac{\epsilon(0)}{4\pi}\phi'(0,t)=D(0,t)\\
                &= \sigma_0^{st}+\frac{1}{2}\big(\sigma^{am}_0(\omega)\exp(\text{i}\omega t)+cc\big).
% \sigma_0^{st}+\hat{\sigma}_0(\omega)\exp(\text{i}\omega t)+cc
  \end{aligned}
  \label{eq:surface_charge_density}
\end{align}
Here we restrict ourselves to express the results in terms of the relative permittivity $\eta$ and
the impedance $Z$. 
The relative permittivity is defined as
\begin{align}
  \eta(\omega)=\frac{\hat{D}(\omega)e/l_B^2}{\epsilon_0\hat{\bar{E}}(\omega)k_BT/(l_Be)}
  =4\pi\frac{\hat{D}(\omega)}{\hat{\bar{E}}(\omega)}=2\pi L\frac{\sigma^{am}_0(\omega)}{\Phi},
  \label{eq:dielectricity}
\end{align}
where $\bar{E}(t)=U(t)/L$ denotes the mean electric field.
The impedance $\hat{U}(\omega)/\hat{I}(\omega)$ reduced by the cross-sectional area $A$ is defined by
\begin{align}
  \frac{\hat{U}(\omega)k_BT/e}{\hat{I}(\omega)e/\tau}Al_B^2
  =\frac{2l_B^4}{e^2\Gamma_{ref}}\frac{\Phi}{\text{i}\omega\sigma^{am}_0(\omega)}=:\frac{2l_B^4}{e^2\Gamma_{ref}}Z(\omega),
  \label{eq:impedance}
\end{align}
where we want to refer to $Z(\omega)$ as the ``reduced impedance''. 
By means of Eqs.~(\ref{eq:dielectricity},\ref{eq:impedance}) one can calculate
the frequency-dependent quantities $\eta(\omega)$ and $Z(\omega)$ from $\sigma^{am}_0(\omega)$ which is 
determined by the electrostatic potential $\phi$ in the vicinity of the electrode at $z=0$ (see Eq.~(\ref{eq:surface_charge_density})).
This requires the knowledge of all the ion density profiles everywhere in the cell.

\subsection{Parameters}
In order to reduce the number of coupled differential equations (\ref{eq:Poisson_equation},
\ref{eq:Nernst_Planck_equation}) we restrict the following discussion to two ion species, 
$\alpha\in\{+,-\}$, with the valencies $q_\pm=\pm1$.

The profiles $\epsilon(z)$ and $V_\pm(z)$, the latter of which describes the spatially varying solubility of
ions, exhibit discontinuities at $z=R$ and $z=S$, respectively:
\begin{align}
  \begin{aligned}
    \epsilon(z)=
    \begin{cases}
      \epsilon_L&, z<R\\
      \epsilon_R&, z>R
    \end{cases},
  \end{aligned}
  \label{eq:solvent_dielectricity}
\end{align}
\begin{align}
  \begin{aligned}
    V_\pm(z)=
    \begin{cases}
      0&, z<S\\
      f_\pm&, z>S
    \end{cases}.
  \end{aligned}
  \label{eq:external_potential}
\end{align}
In general $R\neq S$ because the solubilities of ions deviate from that of the bulk already some
distance away from the dielectric interface at $z=R$ due to a finite size of solvation shells.
However, the thickness of solvation shells is a few solvent particle diameters, which is typically
much smaller than the size of the cell: $|R-S|\ll L$.
It is possible to have different mobilities $\Gamma_\pm(z)$ in the left and right partial cell. 
We want to interpolate linearly in between $R$ and $S$ to take into account that a hydrated ion may
sense the properties of both partial cells at once:
\begin{align}
  \begin{aligned}
    \Gamma_\pm(z)=
    \begin{cases}
      \Gamma^L_\pm & , z<\text{min}(R,S)\\
      \Gamma^R_\pm & , z>\text{max}(R,S)\\
      \text{linear interpolation} & , \text{in between}
    \end{cases}.
  \end{aligned}
  \label{eq:mobility}
\end{align}

\begin{figure}[!t]
  \includegraphics[width=0.4\textwidth]{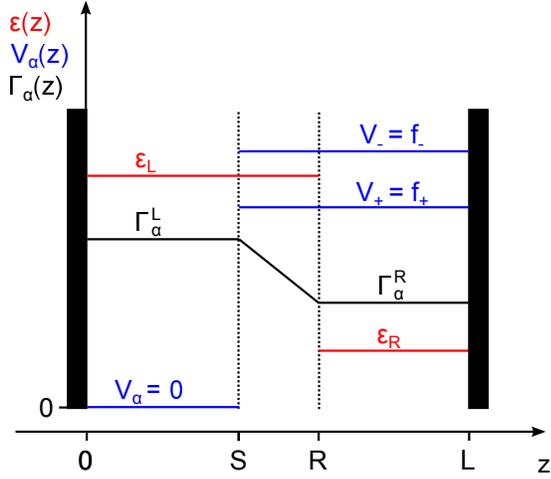}
  \caption{(Color online) Illustration of the profiles in the cell: relative permittivity of the solvents 
           $\epsilon(z)$ with a discontinuity at $z=R$, solubilities of the ions $V_\pm(z)$ with 
           discontinuities at $z=S$ and mobilities of the ions $\Gamma_\pm(z)$ which are linearly 
           interpolated between $z=R$ and $z=S$ in our model.}
  \label{fig:profiles}
\end{figure}

Figure~\ref{fig:profiles} summarizes the parametrization of Eqs. (\ref{eq:solvent_dielectricity}-\ref{eq:mobility}).

%-------------------------------------------------------------------------------

\section{\label{sec:Discussion}Discussion}
\subsection{\label{subsec:Direct_current_voltage}Direct current voltage}
As a first step we want to consider a voltage $U$ that is constant in time. Mathematically this implies time-independent boundary conditions
\begin{align}
  \begin{aligned}
    \phi_0(t)=-\phi_L(t)=\Phi^{st},\\
    j_\pm(0,t)=j_\pm(L,t)=0.
  \end{aligned}
  \label{eq:boundary_conditions_static}
\end{align}
Poisson's equation~(\ref{eq:Poisson_equation}) is a linear differential equation. But as its inhomogeneity (the charge density) is in general not known in analytical form, the equation has to be solved numerically. For that purpose we decide to use a Green's function approach.

The homogeneous equation
\begin{align}
  (\epsilon(z)\phi_{hom}'(z))'=0
\end{align}
can be solved analytically as the coefficients are constant in space in the partial cells:
\begin{align}
  \begin{aligned}
    \phi_{hom}(z,t)=
    \begin{cases}
      \frac{A_L}{\epsilon_L}z+B_L&z<R\\
      \frac{A_R}{\epsilon_R}z+B_R&z>R
    \end{cases}
  \end{aligned}
  \label{eq:linear_system}
\end{align}
In addition to the two boundary conditions on $\phi$ from Eq.~(\ref{eq:boundary_conditions_static})
two more conditions are required in order to solve the linear system Eq.~(\ref{eq:linear_system}).
Firstly, we demand continuity of the electrostatic potential $\phi$ --- also at $z=R$. Secondly, continuity of the dielectric displacement $D$ is assumed which implies that there is no \emph{surface} charge at the interface. 
Hence the two additional conditions are
\begin{align}
  \begin{aligned}
    \phi(R^-,t)&=\phi(R^+,t)\\
    D(R^-,t)=D(R^+,t)&\Leftrightarrow \epsilon_L\phi'(R^-,t)=\epsilon_R\phi'(R^+,t).
  \end{aligned}
  \label{eq:interface_conditions}
\end{align}
With the conditions Eqs. (\ref{eq:boundary_conditions_static},\ref{eq:interface_conditions}) the linear system
Eq.~(\ref{eq:linear_system}) is determined and the solution reads
\begin{align}
 \begin{aligned}
   \phi_{hom}(z,t)=
   \begin{cases}
    -\frac{2\Phi^{st}}{\epsilon_L}\left(\frac{R}{\epsilon_L}+\frac{L-R}{\epsilon_R}\right)^{-1}z+\Phi^{st}&\!z<R\\
    \frac{2\Phi^{st}}{\epsilon_R}\left(\frac{R}{\epsilon_L}+\frac{L-R}{\epsilon_R}\right)^{-1}\!\!(L-z)-\Phi^{st}&\!z>R
   \end{cases}
 \end{aligned}
 \label{eq:phi_homogeneous}
\end{align}
A particular solution of Poisson's equation~(\ref{eq:Poisson_equation})
is obtained by using the Green's function $G(z,z')$ which is defined by 
\begin{align}
  \quad(\epsilon(z)G'(z,z'))'=-4\pi\delta(z-z').
  \label{eq:Poissons_equation_Green}
\end{align}
Since the homogeneous solution $\phi_{hom}$ fulfills the Dirichlet boundary conditions Eq.~(\ref{eq:boundary_conditions_static}), the Green's function $G$ is required to fulfill homogeneous Dirichlet boundary conditions 
and continuity conditions:
\begin{align}
  \begin{aligned}
    G(0,z')&=0\\
    G(L,z')&=0\\
    G(R^-,z')&=G(R^+,z')\\
    \epsilon_L G'(R^-,z')&=\epsilon_R G'(R^+,z').
  \end{aligned}
\end{align}
With this the Green's function is given by
\begin{align}
  \begin{aligned}
    G(z,z')&=
    \begin{cases}
      -\frac{4\pi}{\epsilon_L}\Theta(z-z')(z-z')-\frac{C(z')}{\epsilon_L}z &, z<R \\
      -\frac{4\pi}{\epsilon_R}\Theta(z-z')(z-z')-\frac{C(z')}{\epsilon_R}z &, z>R \\
      \qquad+\frac{4\pi}{\epsilon_R}(L-z')+\frac{L}{\epsilon_R}C(z')
    \end{cases},\\
    C(z')&=\frac{4\pi}{\frac{R-L}{\epsilon_R}-\frac{R}{\epsilon_L}}\bigg[\Theta(R-z')(R-z')\left(\frac{1}{\epsilon_L}-\frac{1}{\epsilon_R}\right)\\
                         &\phantom{=}+(L-z')\frac{1}{\epsilon_R}\bigg].
  \end{aligned}
  \label{eq:Greens_function}
\end{align}
Finally we can write down the full solution of Poisson's equation (\ref{eq:Poisson_equation}) as
\begin{align}
  \phi(z,t)=\phi_{hom}(z,t)+\int\limits_0^L dz'G(z,z')\left(\varrho_+(z',t)-\varrho_-(z',t)\right).
  \label{eq:Poissons_equation_solution}
\end{align}
The time evolution of the ion densities is given by Eq.~(\ref{eq:Nernst_Planck_equation}). Every initial state $\varrho_\pm(z,t=0)$ relaxes for long times to $\varrho_\pm^{st}(z)=\lim\limits_{t\rightarrow\infty}\varrho_\pm(z,t)$. The ultimate chemical potentials $\mu_\pm^{st}$ are constant in time and space, i.e., $\varrho_\pm^{st}(z)$ is the equilibrium state for the static voltage $U=2\Phi^{st}$.

\begin{figure}[!t]
  \includegraphics[width=0.45\textwidth]{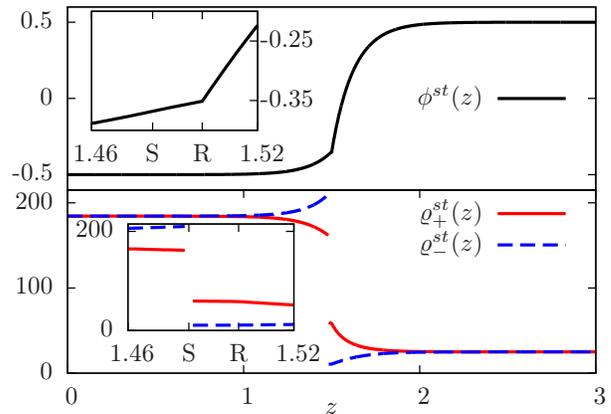}
  \caption{(Color online) Spatial dependence of the static ion densities $\varrho^{st}_\pm(z)$ and of the electrostatic potential $\phi^{st}(z)$. The parameters were chosen as follows: $L=3$, $R=1.5$, $S=1.482$, $\Phi^{st}=-0.5$, $\epsilon_L=80$, $\epsilon_R=10$, $f_+=1$, $f_-=3$, $\bar{\varrho}=104$. (See main text for further information.)}
  \label{fig:profiles_static}
\end{figure}

In Fig.~\ref{fig:profiles_static} the spatial dependence of $\varrho^{st}_\pm(z)$ and $\phi^{st}(z)$ is shown for one particular choice of parameters. Because of $\epsilon_L>\epsilon_R$, the electric field $|\partial_z\phi^{st}(z)|$ is largest in the range $z>R$. The inset shows that the electrostatic potential exhibits a kink at $z=R$ which is the location where $\epsilon(z)$ has its discontinuity. As $f_->f_+$ the negatively charged ions are dissolved less in the range $z>S$ than the positively charged ions. This implies a positive net charge in the right partial cell and, as we assume that the cell is charge neutral, this leads to a negative net charge in the left partial cell. As opposite charges attract each other there is an electric double layer at the interface. The influence of the interface decreases with increasing distance which is why the densities of positive and negative ions have the same asymptotic behavior. The limits can be determined 
analytically. If we neglect any surface contributions the expression of the chemical potential Eq.~(\ref{eq:chemical_potential}) allows one to derive the following estimates for the ionic strengths $\bar{\varrho}_{L,R}$:
\begin{align}
 \begin{aligned}
  \bar{\varrho}_L&=\bar{\varrho}L\left[S+(L-S)\cdot\exp\left(-\frac{f_++f_-}{2}\right)\right]^{-1}\stackrel{Fig.~\ref{fig:profiles_static}}\approx185\\
  \bar{\varrho}_R&=\bar{\varrho}L\left[L-S+S\cdot\exp\left(\frac{f_++f_-}{2}\right)\right]^{-1}\stackrel{Fig.~\ref{fig:profiles_static}}\approx25
 \end{aligned}
 \label{eq:ionic_strengths}
\end{align}
with the average ion density $\bar{\varrho}$. For the cases $f_+=f_-$ and $\Phi^{st}=0$ the static profiles $\varrho_\pm^{st}(z)$ are constants in the partial cells and the respective values are given by Eq.~(\ref{eq:ionic_strengths}).\\
In order to save space we do not show static profiles for the case when $R=S$, i.e., the discontinuities are located
at the same position. Actually the curves look very similar to the ones in Fig.~\ref{fig:profiles_static} (when the insets are ignored). The influence of the interface extent seems to be negligible.\\
If we choose $R=S$ and assume the electrodes to be electrically isolated from the outside the profile of the electric potential $\phi^{st}(z)$ 
coincides with that obtained within
the Verwey-Niessen approach (see Ref. \cite{Verwey1939}).

\subsection{\label{subsec:Alternating_current_voltage}Alternating current voltage}
Now we want to consider the case when the voltage source applies potential differences which oscillate harmonically in time:
\begin{align}
  \begin{aligned}
    \phi_0(t)=-\phi_L(t)=\Phi^{st}+\frac{1}{2}\big(\Phi\exp(\text{i}\omega t)+cc\big),\\
    j_\pm(0,t)=j_\pm(L,t)=0.
  \end{aligned}
  \label{eq:boundary_conditions_harmonic}
\end{align}

\subsubsection{\label{subsec:Greens_function}Green's function}
The direct numerical solution of Eq.~(\ref{eq:Nernst_Planck_equation}) by solving Poisson's equation (\ref{eq:Poisson_equation}) with a Green's function similar to the one in Sec. \ref{subsec:Direct_current_voltage} is subject to certain restrictions:
\begin{itemize}
 \item In order to suppress higher harmonics and to reduce errors when determining the linear response quantities $Z$ and $\eta$ a sufficiently small voltage amplitude $\Phi\ll1$ has to be applied.

 \item The algorithm is very inefficient.
\end{itemize}

\subsubsection{Linearized equations}
Since the reduced impedance $Z(\omega)$ and the relative permittivity $\eta(\omega)$ describe the linear response of the electrolytic cell
to a small voltage $\Phi$, it is sufficient to consider the linearization of Eqs. (\ref{eq:chemical_potential},\ref{eq:Ficks_law},\ref{eq:continuity_equation},\ref{eq:Nernst_Planck_equation}) around the static profiles (superscript $^{st}$). By means of the ansatz
\begin{align}
 \begin{aligned}
  \varrho_\pm(z,t)=&\varrho_\pm^{st}(z)+\frac{1}{2}\big(\varrho_\pm^{am}(z)\exp(\text{i}\omega t)+cc\big),\\
                    &\text{with}\quad\frac{|\varrho_\pm^{am}(z)|}{|\varrho_\pm^{st}(z)|}\ll 1,\\
  \phi(z,t)=&\phi^{st}(z)+\frac{1}{2}\big(\phi^{am}(z)\exp(\text{i}\omega t)+cc\big),\\
                    &\text{with}\quad\phi^{st}(0)=\Phi^{st},\quad\phi^{am}(0)=\Phi
 \end{aligned}
\end{align}
and by neglecting terms of higher than linear order in the amplitude parts (superscript $^{am}$) one obtains
\begin{align}
  \label{eq:chemical_potential_linear}
  \mu_\pm^{am}(z)&=\frac{\varrho_\pm^{am}(z)}{\varrho_\pm^{st}(z)}\pm\phi^{am}(z),\\
  \label{eq:current_density_linear}
  j^{am}_\pm(z)&=-\Gamma_\pm(z)\varrho_\pm^{st}(z)\left(\frac{\partial}{\partial z}\frac{\varrho_\pm^{am}(z)}{\varrho_\pm^{st}(z)}\pm\frac{\partial\phi^{am}(z)}{\partial z}\right),\\
  \label{eq:continuity_equation_linear}
  \text{i}\omega\varrho_\pm^{am}(z)&=-\frac{\partial j_\pm^{am}}{\partial z}(z),\\
  \label{eq:Nernst_Planck_equation_linear}
  \text{i}\omega\varrho_\pm^{am}(z)&=\frac{\partial}{\partial z}\left[\Gamma_\pm(z)\varrho_\pm^{st}(z)\left(\frac{\partial}{\partial z}\frac{\varrho_\pm^{am}(z)}{\varrho_\pm^{st}(z)}\pm\frac{\partial\phi^{am}}{\partial z}\right)\right].
\end{align}
As Poisson's equation (\ref{eq:Poisson_equation}) is a linear differential equation, it holds unchanged for the amplitude parts:
\begin{align}
  \left(\epsilon(z)\frac{\partial\phi^{am}(z)}{\partial z}\right)'=-4\pi(\varrho^{am}_+(z)-\varrho^{am}_-(z)).
  \label{eq:Poissons_equation_am}
\end{align}
Given the static profiles $\varrho_\pm^{st}(z)$, whose time-consuming calculation has to be performed only once for each set of cell parameters, the solution of the linear Eqs. (\ref{eq:chemical_potential_linear}-\ref{eq:Poissons_equation_am}) for each frequency $\omega$ can be determined efficiently.

\subsubsection{\label{subsubsec:The_simple_homogeneous_cell}The simple homogeneous cell}
Equation~(\ref{eq:Nernst_Planck_equation_linear}) is a linear differential equation in the amplitude profiles but there remains a difficulty: 
In general, the static profiles $\varrho_\pm^{st}(z)$ are not available in closed form, which precludes analytic solutions of 
Eq.~(\ref{eq:Nernst_Planck_equation_linear}).
However, in the absence of interfaces inside the cell, i.e., for
\begin{align}
  \begin{aligned}
    \epsilon(z)&=\epsilon,\\
    V_\pm(z)&=0,\\
    \Phi^{st}&=0,
  \end{aligned}
  \label{eq:parameters_simple_homogeneous_cell}
\end{align}
the static ion densities are constant: $\varrho^{st}(z)=\bar{\varrho}$. Such a cell will be called \emph{homogeneous} in the following. For the special case that both ion species have the same mobility, $\Gamma_\pm(z)=\Gamma$, we want to call it a \emph{simple} homogeneous cell.\\
Due to the homogeneity of the cell the system of differential equations (\ref{eq:Nernst_Planck_equation_linear},\ref{eq:Poissons_equation_am}) decouples by considering the sum $\Sigma(z)=\varrho^{am}_+(z)+\varrho^{am}_-(z)$ and the difference $\Delta(z)=\varrho^{am}_+(z)-\varrho^{am}_-(z)$ of the amplitude ion densities. We end up with two equations
\begin{align}
  \label{eq:Nernst_Planck_Sum_homogeneous}
  \Sigma''(z)&=\frac{\text{i}\omega}{\Gamma}\Sigma(z),\\
  \label{eq:Nernst_Planck_Difference_homogeneous}
  \Delta''(z)&=\left(\frac{\text{i}\omega}{\Gamma}+\frac{8\pi \bar{\varrho}}{\epsilon}\right)\Delta(z)=:\zeta^2\Delta(z).
\end{align}
In Eq.~(\ref{eq:Nernst_Planck_Difference_homogeneous}) one can identify $\kappa^2=8\pi \bar{\varrho}/\epsilon$ as the square of the inverse Debye length (see Refs. \cite{Debye1923,McQuarrie2000}). Equations (\ref{eq:Nernst_Planck_Sum_homogeneous},\ref{eq:Nernst_Planck_Difference_homogeneous}) are linear and homogeneous differential equations of second order and may be solved by an exponential ansatz
\begin{align}
  \label{eq:Sum_homogeneous}
  \Sigma(z)&=A\,\text{exp}\left(\sqrt{\frac{\text{i}\omega}{\Gamma}}z\right)+B\,\text{exp}\left(-\sqrt{\frac{\text{i}\omega}{\Gamma}}z\right)\\
  \label{eq:Difference_homogeneous}
  \Delta(z)&=C\,\text{e}^{\zeta z}+D\,\text{e}^{-\zeta z}
\end{align}
with integration constants $A,B,C$ and $D$.\\
Using Eq.~(\ref{eq:Difference_homogeneous}) on the right-hand side of Eq.~(\ref{eq:Poissons_equation_am}) and integrating twice with respect to $z$ one obtains $\phi^{am}(z)$ in terms of $C,D$ and two additional integration constants. The in total six integration constants are uniquely determined by the six boundary conditions Eqs. (\ref{eq:boundary_conditions_harmonic}).\\
With the knowledge of the amplitude profile $\phi^{am}(z)$ of the electrostatic potential the reduced impedance $Z$ and the relative permittivity $\eta$ are given by (see Eqs. (\ref{eq:surface_charge_density}-\ref{eq:impedance}))
\begin{align}
  \label{eq:Dielectricity_homogeneous}
  \eta(\omega)&=\frac{L\epsilon\zeta}{\bar{\varrho}}\left[\frac{16\pi}{\epsilon\zeta^2}\tanh\left(\frac{\zeta}{2}L\right)+\frac{L}{\bar{\varrho}}\left(\zeta-\frac{8\pi \bar{\varrho}}{\epsilon\zeta}\right)\right]^{-1},\\
  \label{eq:Impedance_homogeneous}
  Z(\omega)&=\frac{2\pi \bar{\varrho}}{\epsilon\zeta}\frac{1}{\text{i}\omega}\left[\frac{16\pi}{\epsilon\zeta^2}\tanh\left(\frac{\zeta}{2}L\right)+\frac{L}{\bar{\varrho}}\left(\zeta-\frac{8\pi \bar{\varrho}}{\epsilon\zeta}\right)\right].
\end{align}
A similar calculation shows that for the one-component plasma, where one of the mobilities $\Gamma_\pm$ vanishes,
$Z$ and $\eta$ are obtained from Eqs. (\ref{eq:Dielectricity_homogeneous}) and (\ref{eq:Impedance_homogeneous}) by replacing the ionic strength $\bar{\varrho}$ with half of its value $\bar{\varrho}/2$.

\subsubsection{\label{subsubsec:The_general_homogeneous_cell}The general homogeneous cell}
If the cell is homogeneous, i.e., without a liquid-liquid interface (see Eq.~(\ref{eq:parameters_simple_homogeneous_cell})),
but the ion species have different mobilities, $\Gamma_+\not=\Gamma_-$,
Eqs.~(\ref{eq:Nernst_Planck_equation_linear}) and (\ref{eq:Poissons_equation_am}) do not decouple by introducing $\Sigma(z)$ and $\Delta(z)$.\\
Instead, by inserting Poisson's equation (\ref{eq:Poissons_equation_am}) into the Nernst-Planck equation (\ref{eq:Nernst_Planck_equation_linear}) one obtains
\begin{align}
  \frac{\partial^2\varrho^{am}_\pm(z)}{\partial z^2}=\left(\frac{\text{i}\omega}{\Gamma_\pm}+\frac{4\pi \bar{\varrho}}{\epsilon}\right)\varrho^{am}_\pm(z)-\frac{4\pi \bar{\varrho}}{\epsilon}\varrho^{am}_\mp(z)
  \label{eq:Poisson_in_Nernst_Planck_different_Gammas}
\end{align}
which may be expressed as
\begin{align}
 \begin{aligned}
 \begin{pmatrix}
  \varrho^{am}_+(z)\\
  \varrho^{am}_-(z)
 \end{pmatrix}''=
 \underbrace
   {
  \frac{4\pi \bar{\varrho}}{\epsilon}
   \begin{pmatrix}
     1+a & -1\\
     -1 & 1+b       
   \end{pmatrix}
   }_{=:M}
 \begin{pmatrix}
  \varrho^{am}_+(z)\\
  \varrho^{am}_-(z)
 \end{pmatrix}\\
 \end{aligned}
 \label{eq:matrix_expression}
\end{align}
where $a=\frac{\epsilon}{4\pi I\Gamma_+}\text{i}\omega$ and $b=\frac{\epsilon}{4\pi I\Gamma_-}\text{i}\omega$.
The eigenvalues $\lambda_\pm$ and eigenvectors $v_\pm$ of matrix $M$ are
\begin{align}
 \lambda_\pm=\frac{4\pi \bar{\varrho}}{\epsilon}\left[1+\frac{a+b}{2}\pm\sqrt{1+\left(\frac{a-b}{2}\right)^2}\right]
\end{align}
and
\begin{align}
 \begin{aligned}
 v_\pm=
 \begin{pmatrix}
  1\\
  \frac{a-b}{2}\mp\sqrt{1+\left(\frac{a-b}{2}\right)^2}
 \end{pmatrix}
 =
 \begin{pmatrix}
  1\\
  -\frac{\epsilon}{4\pi \bar{\varrho}}\lambda_\pm+1+a
 \end{pmatrix},
 \end{aligned}
\end{align}
respectively. As the eigenvectors are orthogonal $\left(^tv_\pm\cdot v_\mp=0\right)$, we may express the ion densities in terms of them:
\begin{align}
  \begin{pmatrix}
    \varrho^{am}_+(z)\\
    \varrho^{am}_-(z)
  \end{pmatrix}
  =A(z)v_++B(z)v_-
  \label{eq:relation_densities_eigenvectors}
\end{align}
Equation~(\ref{eq:matrix_expression}) reads
\begin{align}
 \begin{aligned}
 A''(z)=\lambda_+A(z)\\
 B''(z)=\lambda_-B(z)
 \end{aligned}
\end{align}
with the solutions
\begin{align}
 \begin{aligned}
  A(z)=A_1\exp\left(\sqrt{\lambda_+}z\right)+A_2\exp\left(-\sqrt{\lambda_+}z\right)\\
  B(z)=B_1\exp\left(\sqrt{\lambda_-}z\right)+B_2\exp\left(-\sqrt{\lambda_-}z\right)
 \end{aligned}
 \label{eq:conditional_equations_coefficients}
\end{align}
Similarly to Sec. \ref{subsubsec:The_simple_homogeneous_cell} one can integrate Poisson's Eq.~(\ref{eq:Poissons_equation_am})
to obtain $\phi^{am}(z)$ and $\varrho^{am}_\pm(z)$ in terms of six integration constants,
which are determined by the six boundary conditions Eq.~(\ref{eq:boundary_conditions_harmonic}).\\
The system of linear equations is analytically solvable, but the final expressions are rather complex, so that we refrain from showing them here.

\subsubsection{\label{subsubsec:Cell_with_interface_equal_partitioning}Cell with interface --- equal partitioning}
A simple example for a cell with an interface which allows for analytical solutions is given by the following choice of parameters:
\begin{align}
  \begin{aligned}
    R&=S\\
    \epsilon(z)&=
    \begin{cases}
      \epsilon_L &z<R\\
      \epsilon_R &z>R
    \end{cases}\\
    f_+&=f_-=:f\\
    V_\pm(z)&=
    \begin{cases}
      0 &z<R\\
      f &z>R
    \end{cases}\\
    \Gamma_\pm(z)&=
    \begin{cases}
      \Gamma^L &z<R\\
      \Gamma^R &z>R
    \end{cases}\\
    \Phi^{st}&=0
  \end{aligned}
\end{align}
The dielectric interface at $z=R$ and the solubility interface at $z=S$ coincide and the ion species within the partial cells have the same mobilities. As the ion solubility $f_\pm=f$ is the same for both species there will be no net charge in the partial cells. This is why this case is called \emph{equal partitioning}. Furthermore within the partial cells the ion densities will be constants where the values are given by Eq.~(\ref{eq:ionic_strengths}):
\begin{align}
  \begin{aligned}
    \varrho_\pm^{st}(z)=
    \begin{cases}
      \bar{\varrho}_L &z<R\\
      \bar{\varrho}_R &z>R
    \end{cases}.
  \end{aligned}
\end{align}
The profiles of the partial cells equal the ones of the simple homogeneous cell. By introducing, in analogy to Sec. \ref{subsubsec:The_simple_homogeneous_cell}, profiles $\Sigma_{L,R}(z)$ and $\Delta_{L,R}(z)$ for the sums and the differences of the ion densities in the partial cells, one obtains the analogs of Eqs. (\ref{eq:Nernst_Planck_Sum_homogeneous}) and (\ref{eq:Nernst_Planck_Difference_homogeneous}):
\begin{align}
  \begin{aligned}
    &z<R:\\
    &\Sigma_L''(z)=\frac{\text{i}\omega}{\Gamma^L}\Sigma_L(z)\\
    &\Delta_L''(z)=\left(\frac{\text{i}\omega}{\Gamma^L}+\frac{8\pi \bar{\varrho}_L}{\epsilon_L}\right)\Delta_L(z)=:\zeta^2_L\Delta_L(z)\\
    &z>R:\\
    &\Sigma_R''(z)=\frac{\text{i}\omega}{\Gamma^R}\Sigma_R(z)\\
    &\Delta_R''(z)=\left(\frac{\text{i}\omega}{\Gamma^R}+\frac{8\pi \bar{\varrho}_R}{\epsilon_R}\right)\Delta_R(z)=:\zeta^2_R\Delta_R(z)
  \end{aligned}
  \label{eq:equal_partitioning_decoupled_Poisson_Nernst_Planck}
\end{align}
Again, integrating Poisson's Eq.~(\ref{eq:Poissons_equation_am}) twice in each partial cell leads to six integration constants per partial cell, i.e., to twelve integration constants in total. Since Eq.~(\ref{eq:boundary_conditions_harmonic}) provides only six boundary conditions at the electrodes, one has to require six additional boundary conditions at the interface:
\begin{itemize}
  \item electrostatic potential:
        \begin{align}
          \phi^{am}(R^-)=\phi^{am}(R^+)
          \label{eq:continuity_phi_am}
        \end{align}
  \item chemical potential
        \begin{align}
          \mu_\pm^{am}(R^-)=\mu_\pm^{am}(R^+)
          \label{eq:continuity_mu_am}
        \end{align}
        (Otherwise the current densities $j^{am}_\pm$ would exhibit an unphysical $\delta$-singularity at $z=R$ (see Eq.~(\ref{eq:current_density_linear})).)
  \item current densities
        \begin{align}
          j^{am}_\pm(R^-)&=j^{am}_\pm(R^+)
          \label{eq:continuity_j_am}
        \end{align}
        (Otherwise the ion densities $\varrho^{am}_\pm$ at $z=R$ would change infinitely fast (see Eq.~(\ref{eq:continuity_equation_linear})), which is unphysical.)
  \item dielectric displacement
        \begin{align}
          \begin{aligned}
            D^{am}(R^-)&=D^{am}(R^+)\\
            \Leftrightarrow \epsilon(R^-)\frac{\partial\phi^{am}}{\partial z}(R^-)&=\epsilon(R^+)\frac{\partial\phi^{am}}{\partial z}(R^+)
          \end{aligned}
          \label{eq:continuity_D_am}
        \end{align}
\end{itemize}
Again the final analytic solution is too complex to be shown here.

\subsubsection{\label{subsubsec:Cell_with_interface_unequal_partitioning}Cell with interface --- unequal partitioning}
\begin{figure*}[t]
  \includegraphics[width=0.95\textwidth]{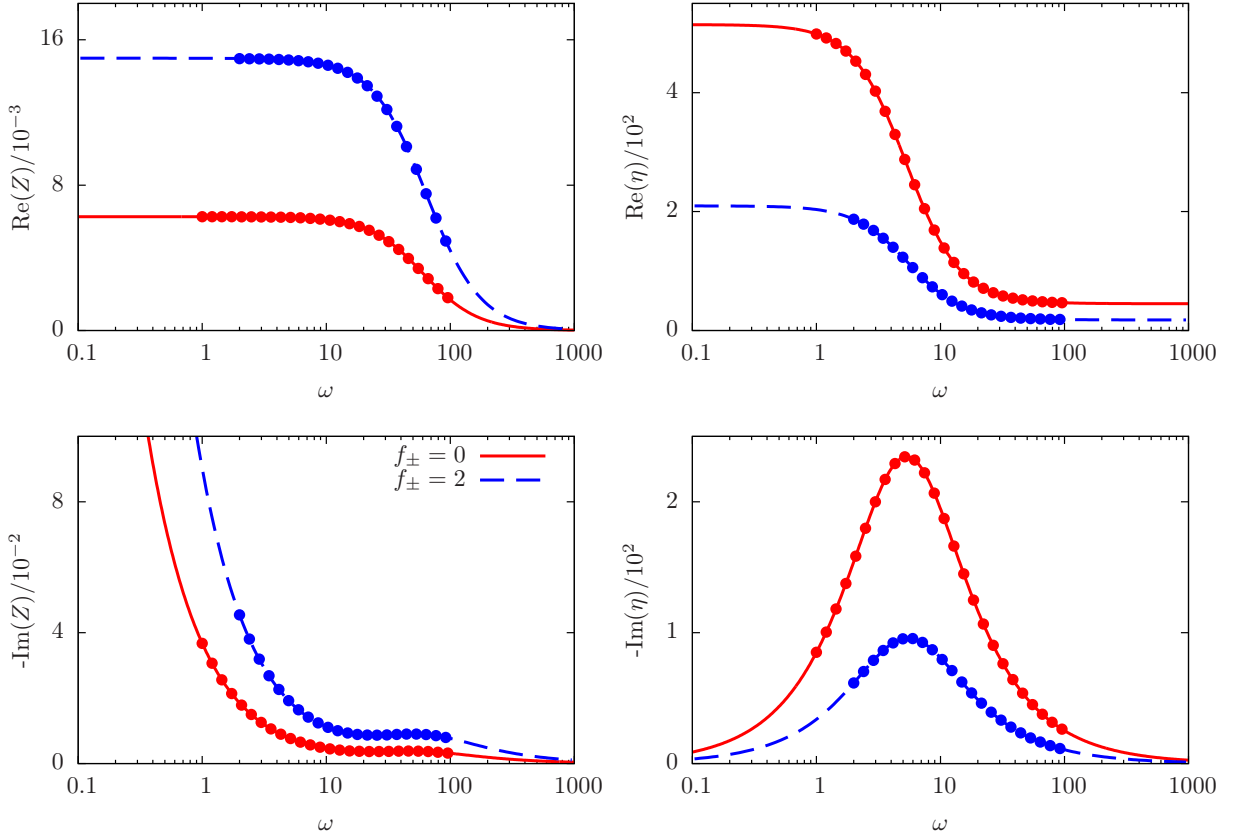}
  \caption{(Color online) Comparison of reduced impedance $Z(\omega)$ and relative permittivity $\eta(\omega)$ spectra of simple cells with (dashed blue lines) and without (solid red lines) an interface. No qualitative differences are visible for the two cells. The lines are results of the approaches in Secs. \ref{subsubsec:The_simple_homogeneous_cell} and \ref{subsubsec:Cell_with_interface_equal_partitioning}, respectively, whereas the dots have been obtained by the direct approach of Sec. \ref{subsec:Greens_function}. The common parameters are $L=3$, $\Phi^{st}=0$, $\bar{\varrho}=104$, $\Gamma_\pm=1$, $\Phi=0.01$. The solid red curves correspond to $f_\pm=0$, $\epsilon=45$, whereas the dashed blue curves correspond to $f_\pm=2$, $\epsilon_L=80$, $\epsilon_R=10$, $R=S=L/2$.}
  \label{fig:spectra_with_and_without_interface}
\end{figure*}
As soon as the solubilities of the ions are unequal ($f_+\neq f_-$) the static ion densities $\varrho^{st}_\pm(z)$ are no longer spatially constant (see, e.g., Fig.~\ref{fig:profiles_static}). The fact that in general no analytical expressions for $\varrho^{st}_\pm(z)$ are available precludes analytical solutions for the amplitude profiles (superscript $^{am}$). Hence one has to rely on numerical methods. In order to calculate the reduced impedance or relative permittivity spectra faster than with the approach of Sec. \ref{subsec:Greens_function} one may benefit from the linearity of the differential Eqs. (\ref{eq:Nernst_Planck_equation_linear}). It is possible to derive a very fast algorithm to find the amplitude profiles numerically in the cases of unequal partitioning. Further detail may be found in Ref. \cite{Reindl2012}.

\subsubsection*{}
The approaches of Secs. \ref{subsubsec:The_simple_homogeneous_cell}-\ref{subsubsec:Cell_with_interface_unequal_partitioning} are based on the linearized Eqs. (\ref{eq:Nernst_Planck_equation_linear}) to calculate the ion dynamics when a small alternating current voltage is applied. The main advantage compared to the brute force approach of Sec. \ref{subsec:Greens_function}, where the problem is solved directly by considering the exact Eq.~(\ref{eq:Nernst_Planck_equation}), is the reduction of computation time by several orders of magnitudes.\\
However, the different approaches offer the possibility to avoid implementation errors. Figure~\ref{fig:spectra_with_and_without_interface} shows two different reduced impedance and relative permittivity spectra for cells with (dashed blue lines) and without (solid red lines) an interface which have been determined by different approaches. Comparison of the solid and the dashed spectra leads to the insight that there is no visible qualitative difference between the cells with and without interfaces. This can be expected since the spectra are dominated by bulk contributions. One has to subtract them in order to obtain interfacial properties. The next section describes a possible method.

\subsection{Equivalent circuits\label{subsec:Equivalent_circuits}}
Equivalent circuits have often been used to analyse measured impedance spectra (see, e.g., Refs. \cite{Samec1981,Hajkova1983,Geblewicz1984,Samec1987,Wandlowski1988,Macdonald1992,Silva2005,Barreira2004,Samec2004,Vanysek2008,Ruan2009,Patricio2010}).
One tries to find an electric circuit whose impedance spectrum is similar to the measured one based on the well-known properties of the circuit elements, e.g.,
Ohmic resistors or capacitors, which are interpreted in terms of certain 
microscopic processes.
In this work we use equivalent circuits to extract 
properties of interfaces of two ion conducting liquids from impedance spectra.
As has been explained in the introduction (see Sec.~\ref{sec:intro}), we are \emph{not} following the
standard approach of attempting to find a more or less complex equivalent circuit with elements corresponding 
to bulk or interfacial processes which is fitted to an impedance spectrum in one step. 
Rather, in order to achieve a unique characterization of an equivalent circuit element as corresponding
to either a bulk or an interfacial process and in order to avoid a possible reduction of errors of the bulk 
elements at the expense of an enhancement of errors of the interfacial
elements, we instead apply the following two-step scheme: We first seek to determine the impedance of 
the bulk of the partial cells by means of according homogeneous cells and afterwards determine the interfacial 
impedance by subtracting the bulk impedances from the total impedance.

In Sec. \ref{subsubsec:equivalent_circuit_for_simple_homogeneous_cells} an equivalent circuit for the simple homogeneous cell is derived from the analytical expression for the reduced impedance Eq.~(\ref{eq:Impedance_homogeneous}). This circuit will be extended in Sec. \ref{subsubsec:equivalent_circuit_for_general_homogeneous_cells} to represent general homogeneous cells. Finally in Sec. \ref{subsubsec:equivalent_circuit_for_simple_cells_with_interface} an electrolytic cell containing an interface is modeled by means of the equivalent circuits of the simple type. In Fig.~\ref{fig:equivalent_circuit_simple_interface_cell} the parallel circuit in the middle represents the interface. We will refer to the Ohmic resistance $R_{pi}$ and the capacitor $C_{pi}$ as the \emph{interface 
elements}.

\subsubsection{\label{subsubsec:equivalent_circuit_for_simple_homogeneous_cells}Equivalent circuit for simple homogeneous cells}
\begin{figure}[!t]
  \includegraphics[width=0.45\textwidth]{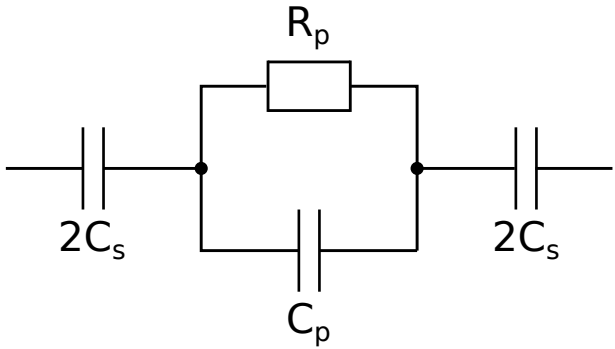}
  \caption{Equivalent circuit for simple homogeneous cells consisting of an Ohmic resistor $R_p$ and two capacitors $C_p$ and $C_s$. The latter represents the electrodes which is why one capacitor $C_s$ has been replaced by a series circuit of two capacitors $2C_s$. In this way they can be positioned in order to visualize their task. Additionally $2C_s$ is the capacitance of a capacitor whose plates are separated by one Debye length $1/\kappa$ which is a common estimate for the range of surface effects in electrolyte solutions. The connection between the circuit elements and the cell properties is given by Eqs. (\ref{eq:components_circuit_simple_homogeneous_cell}). The impedance of the shown circuit is given by Eq.~(\ref{eq:Impedance_circuit_simple_homogeneous}).}
  \label{fig:equivalent_circuit_simple_homogeneous_cell}
\end{figure}

\begin{figure}[!t]
  \includegraphics[width=0.45\textwidth]{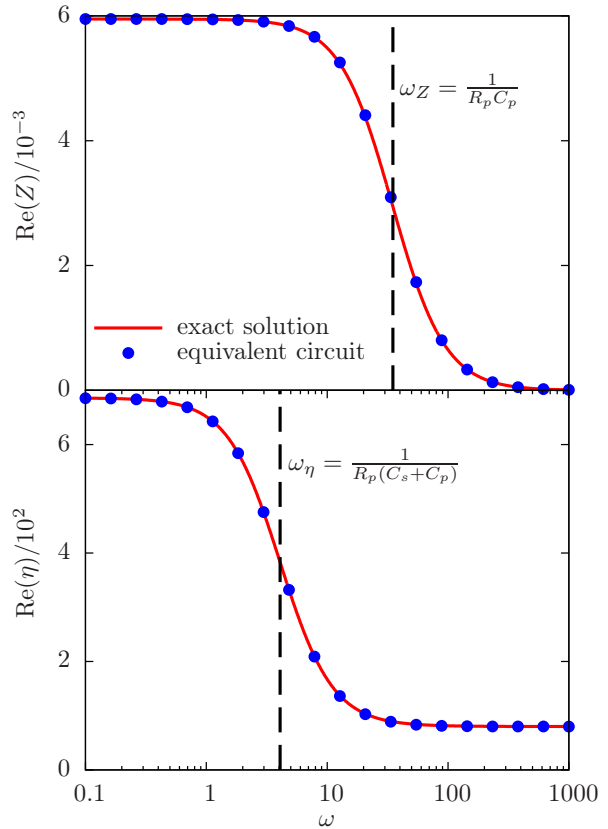}
  \caption{(Color online) Frequency dependence of the real parts of the reduced impedance $Z$ and of the relative permittivity $\eta$ for simple homogeneous cells. The exact solution Eqs. (\ref{eq:Dielectricity_homogeneous}) and (\ref{eq:Impedance_homogeneous}) (solid red lines) and the equivalent circuit approximation Eq.~(\ref{eq:Impedance_circuit_simple_homogeneous}) (blue dots) are indistinguishable for all frequencies $\omega$. The dashed vertical lines refer to frequencies Eqs. (\ref{eq:characteristic_frequency_ReZ_homogeneous},\ref{eq:characteristic_frequency_Reeta_homogeneous}). The parameters for these plots are $L=3$, $\Phi^{st}=0$, $\bar{\varrho}=104$, $\epsilon=80$ and $\Gamma_\pm=1$.}
  \label{fig:spectra_homogeneous_cell_equivalent_circuit}
\end{figure}

\begin{figure}[!t]
  \includegraphics[width=0.45\textwidth]{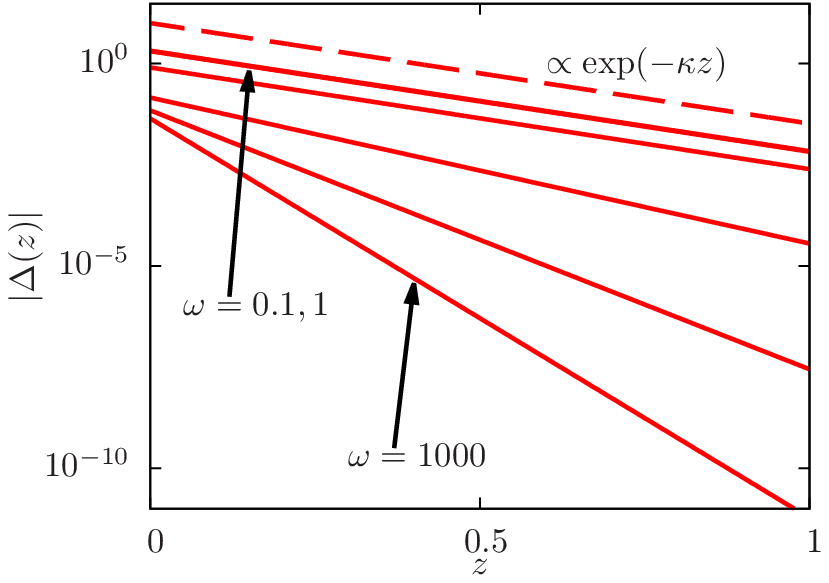}
  \caption{(Color online) Amplitude of the charge density $|\Delta(z)|=|\varrho^{am}_+(z)-\varrho^{am}_-(z)|$ obtained by the method
           of Sec. \ref{subsubsec:The_simple_homogeneous_cell} as a function
           of the location $z$ in the cell for the frequencies $\omega=0.1,1,10,100,400,1000$.
           With increasing frequency both the amplitude at the electrode and the decay length decrease.
           The parameters for these plots are $L=3$, $\Phi^{st}=0$, $\bar{\varrho}=104$, $\epsilon=80$ and $\Gamma_\pm=1$.}
  \label{fig:profiles_harmonic_chargedensity_logarithmic}
\end{figure}

\begin{figure}[!t]
  \includegraphics[width=0.45\textwidth]{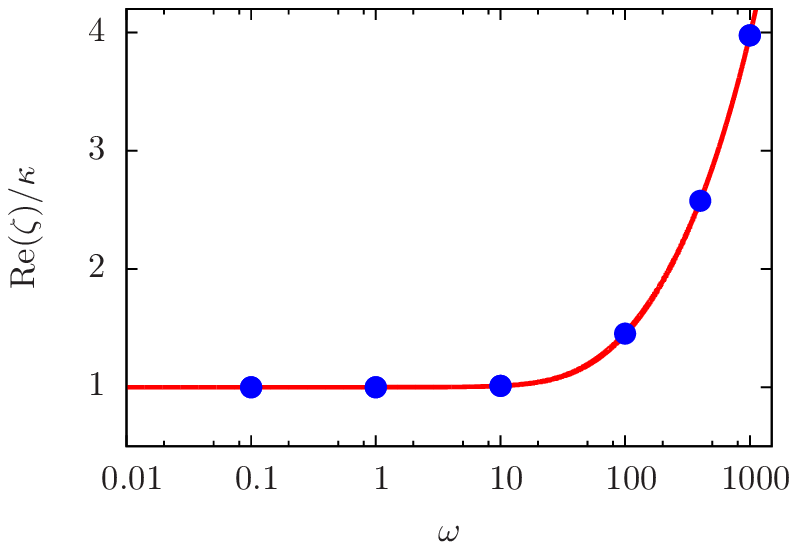}
  \caption{(Color online) Frequency dependence of the real part of the inverse decay length $\Re(\zeta)$ of the charge density $\Delta(z)$ (see Eq.~(\ref{eq:Nernst_Planck_Difference_homogeneous})). For low frequencies it can be estimated by the inverse Debye length $\kappa$. However its value grows with increasing frequency which is the explanation for the decay behavior of the charge density in Fig.~\ref{fig:profiles_harmonic_chargedensity_logarithmic}. The dots correspond to the frequencies which are shown there. The parameters for the plot are $\bar{\varrho}=104$, $\epsilon=80$ and $\Gamma_\pm=1$.}
  \label{fig:zeta_frequencydependent}
\end{figure}

\begin{figure}[!t]
  \includegraphics[width=0.45\textwidth]{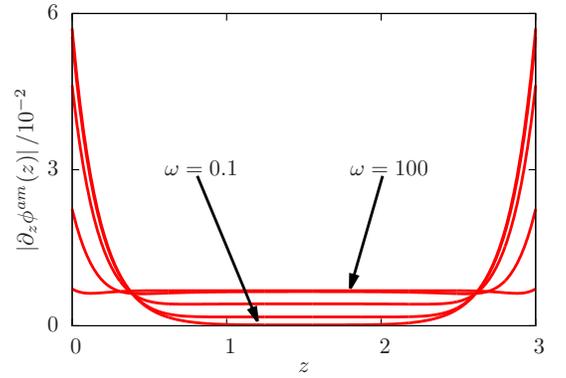}
  \caption{(Color online) Amplitude of the electric field $|\partial_z\phi^{am}(z)|$ obtained by the method
           of Sec. \ref{subsubsec:The_simple_homogeneous_cell} as a function of the location $z$ in the cell for the
           frequencies $\omega=0.1,1,3,10,100$. The value of the electric field in the middle of the cell increases with frequency as the 
           ability of the ions to screen the field decreases with frequency. The parameters for these plots are 
           $L=3$, $\Phi^{st}=0$, $\bar{\varrho}=104$, $\epsilon=80$ and $\Gamma_\pm=1$.}
  \label{fig:profiles_harmonic_electricfield}
\end{figure}

Mathematically the impedance of a circuit consisting of a finite number of Ohmic resistors (impedance independent of frequency) and capacitors (impedance $\sim(\text{i}\omega)^{-1}$) is a rational function of the frequency $\omega$. However, the case of a simple homogeneous cell in Sec. \ref{subsubsec:The_simple_homogeneous_cell} shows a reduced impedance which is transcendental in $\omega$ (see Eq.~(\ref{eq:Impedance_homogeneous})) and it cannot be expected that cells which are more complex should give rise to a simpler, algebraic dependence on $\omega$. This is why all following circuits are never exact, but ideally good approximations (see also Ref. \cite{Macdonald1977}). An equivalent circuit for the simple homogeneous cell can be derived directly from the analytical solution in Eq.~(\ref{eq:Impedance_homogeneous}). Combining the asymptotic forms of the reduced impedance for $\omega\rightarrow 0$ and $\omega\rightarrow \infty$ leads to
\begin{align}
  Z(\omega)=\frac{1}{\text{i}\omega C_s}+\frac{1}{\frac{1}{R_p}+\text{i}\omega C_p}
  \label{eq:Impedance_circuit_simple_homogeneous}
\end{align}
which corresponds to the circuit in Fig.~\ref{fig:equivalent_circuit_simple_homogeneous_cell},
where for $L\to\infty$
\begin{align}
  \begin{aligned}
    R_p&\simeq\frac{L}{4\bar{\varrho}\Gamma},\\
    C_s&\simeq\sqrt{\frac{\epsilon \bar{\varrho}}{2\pi}}=\frac{\epsilon\kappa}{4\pi},\\
    C_p&\simeq\frac{\epsilon}{2\pi L}.
  \end{aligned}
  \label{eq:components_circuit_simple_homogeneous_cell}
\end{align}
This kind of equivalent circuit Eq.~(\ref{eq:Impedance_circuit_simple_homogeneous}) has already been proposed to describe ion transport in a cell with blocking electrodes (see Ref. \cite{Macdonald1977,Dyre2009}). The resulting reduced impedance and relative permittivity spectra agree with the exact solution even at intermediate frequencies (see Fig.~\ref{fig:spectra_homogeneous_cell_equivalent_circuit}).
For reasons of clarity Eqs. (\ref{eq:components_circuit_simple_homogeneous_cell}) display only the asymptotic behavior for large cells ($L\rightarrow\infty$). The lengthy full expressions can be found in Ref. \cite{Reindl2012}.
The asymptotic expression of $R_p$ in Eq. (\ref{eq:components_circuit_simple_homogeneous_cell}) corresponds to Ohm's law of ions with mobility $\Gamma$ moving in a cell of length $L$ driven by a uniform electric field. $C_s$ and $C_p$ are capacitances of capacitors with dielectric $\epsilon$ whose plates are separated by the distance $2/\kappa$ and $L$, respectively. This finding leads to the interpretation that $C_s$ represents the blocking electrodes. The quantity $2C_s$ is 
related to the \emph{double-layer capacity} \cite{Schmickler2010,Chapman1913}.
In Fig.~\ref{fig:equivalent_circuit_simple_homogeneous_cell} one capacitor of capacitance $C_s$ was splitted into two capacitors of capacitance $2C_s$ respectively. In that way each capacitor represents one electrode with an 
effective extent of one Debye length. $R_p$ and $C_p$ however describe the 
bulk processes in
the cell as both components depend on the cell length $L$ (see Eq.~(\ref{eq:components_circuit_simple_homogeneous_cell})).
In Fig.~\ref{fig:profiles_harmonic_chargedensity_logarithmic} the absolute value of the amplitude of the charge density $\Delta(z)$ 
is plotted as a function of the
position $z$ for various frequencies $\omega$.
All of the profiles can be identified as exponential decays as they are 
straight lines
with a negative slope in the semi-logarithmic plot of Fig.~\ref{fig:profiles_harmonic_chargedensity_logarithmic}. This 
observation
is in agreement with Eq.~(\ref{eq:Nernst_Planck_Difference_homogeneous}) where $\zeta=\sqrt{\text{i}\omega/\Gamma+\kappa^2}$ denotes the characteristic complex 
inverse decay length,
whose real part is displayed in Fig.~\ref{fig:zeta_frequencydependent}.
In the quasistatic regime at low frequencies, where the time $2\pi/\omega$ of one period is sufficiently long to form the diffuse layer
in the vicinity of the electrodes, $\Re(\zeta)$ can be estimated by the inverse Debye length $\kappa$.
This is demonstrated
in Fig.~\ref{fig:profiles_harmonic_chargedensity_logarithmic}, where the profiles 
of $|\Delta(z)|$ 
at low frequencies 
decay with the Debye length $1/\kappa$ shown by the dashed line.
This behavior affects 
the electrical field strength $|\partial_z\phi^{am}(z)|$ (see Fig.~\ref{fig:profiles_harmonic_electricfield}),
which at low frequencies almost vanishes
in the middle of the cell because 
of an efficient screening
by the ions. The real part of the reduced impedance 
$\Re(Z(\omega\rightarrow0))\simeq R_p$ can be understood within the picture of 
ions moving in an external field (see Eq.~(\ref{eq:components_circuit_simple_homogeneous_cell})). The real part $\Re(\eta(\omega))$ of the relative permittivity 
is largest at low frequencies (see Fig.~\ref{fig:spectra_homogeneous_cell_equivalent_circuit}), where screening leads to a strongly decaying electric field at the electrode at $z=0$ (see Fig.~\ref{fig:profiles_harmonic_electricfield}), which corresponds to a large surface
charge $\sigma_0$ (see Eq.~(\ref{eq:surface_charge_density})).

At high frequencies the magnitude $|\Delta(z)|$ of the charge density at $z=0$ is smaller than at low frequencies,
because the time $2\pi/\omega$ of one period is not long enough for the ions to completely screen the surface charge.
Moreover the charge density decays on shorter length scales than at low frequencies (see Fig.~\ref{fig:profiles_harmonic_chargedensity_logarithmic}). 
Figure~\ref{fig:zeta_frequencydependent} displays 
the inverse decay length $\Re(\zeta)$, 
which increases $\sim\sqrt{\omega}$ for large frequencies $\omega$
so that Eq.~(\ref{eq:Nernst_Planck_Difference_homogeneous}) takes the form of a diffusion equation. This means that at high frequencies the ions cannot follow the 
rapidly oscillating external field, but the charge transport in this frequency range is dominated by diffusion processes. 
In the equivalent circuit Fig.~\ref{fig:equivalent_circuit_simple_homogeneous_cell} at high frequencies the current flows mainly 
in the lower branch with the capacitor $C_p$, which allows merely for charge oscillations, but which prevents charge transport. 
According to Eq.~(\ref{eq:Impedance_circuit_simple_homogeneous}) the reduced impedance approaches zero for high frequencies (see 
Fig.~\ref{fig:spectra_homogeneous_cell_equivalent_circuit}). At high frequencies the real part of the relative permittivity
approaches the relative permittivity of the solvent, $\Re(\eta(\omega\rightarrow\infty))\simeq\epsilon$,
which means that the ions do not influence the electric field inside the cell. In accordance to this result
one infers from Fig.~\ref{fig:profiles_harmonic_electricfield}
that the electric field becomes homogeneous at high frequencies because the ions cannot screen it any more.

At intermediate frequencies both $\Re(Z)$ and $\Re(\eta)$ exhibit a crossover between their low and high frequency limits (see Fig.~\ref{fig:spectra_homogeneous_cell_equivalent_circuit}).
The crossover in $\Re(Z)$ occurs at the frequency 
\begin{align}
  \omega_{Z}=\frac{1}{R_pC_p}\stackrel{L\to\infty}{\simeq}\kappa^2\Gamma,
  \label{eq:characteristic_frequency_ReZ_homogeneous}
\end{align}
where the real part of the circuit impedance $\Re(Z)$ of Eq.~(\ref{eq:Impedance_circuit_simple_homogeneous}) is half of the value of its low frequency limit. 
Rewriting the limiting form of Eq.~(\ref{eq:characteristic_frequency_ReZ_homogeneous}) as $\kappa^2\simeq\omega_Z/\Gamma$ offers the
intuitive picture of the crossover to occur at the frequency $\omega\approx\omega_Z$, where the ion movement changes from electric field-driven at low frequencies ($\zeta(\omega)^2\approx\kappa^2$, see Eq.~(\ref{eq:Nernst_Planck_Difference_homogeneous})) to diffusive at high frequencies ($\zeta(\omega)^2\approx i\omega/\Gamma$).
In Fig.~\ref{fig:spectra_homogeneous_cell_equivalent_circuit} the frequency $\omega_Z$ is shown by a vertical dashed line which is close to the inflection point. The estimate for $L\rightarrow\infty$ in Eq.~(\ref{eq:characteristic_frequency_ReZ_homogeneous}) reveals that $\omega_Z$ is independent of the cell length $L$. In case of the relative permittivity the crossover in $\Re(\eta)$ can be understood as boundary between the ranges in which the ions effectively screen the external field (lower frequencies) and in which the ions cannot screen the field any more (higher frequencies). Actually there is a correlation between the value of $\Re(\eta)$ in Fig.~\ref{fig:spectra_homogeneous_cell_equivalent_circuit} and the value of the electric field $|\partial_z\phi^{am}(z)|$ in the middle of the cell (see Fig.~\ref{fig:profiles_harmonic_electricfield}): a high 
$\Re(\eta)$ corresponds to a low $|\partial_z\phi^{am}(z)|$. Again the frequency of the crossover may be estimated with help of the equivalent circuit elements:
\begin{align}
  \omega_{\eta}=\frac{1}{R_p(C_s+C_p)}\stackrel{L\to\infty}{\simeq}\frac{\kappa\Gamma}{L/2}
  \label{eq:characteristic_frequency_Reeta_homogeneous}
\end{align}
In Fig.~\ref{fig:spectra_homogeneous_cell_equivalent_circuit} the frequency $\omega_\eta$ is shown by a vertical dashed line which is close to the inflection point. 
$\omega_\eta$, in contrast to $\omega_Z$, exhibits a dependency on the cell length $L$, since the driving force, which is proportional
to the electric field $2\Phi/L$ inside the cell, weakens and hence screening slows down upon increasing $L$.

\subsubsection{\label{subsubsec:equivalent_circuit_for_general_homogeneous_cells}Equivalent circuit for general homogeneous cells}
\begin{figure}[!t]
  \includegraphics[width=0.45\textwidth]{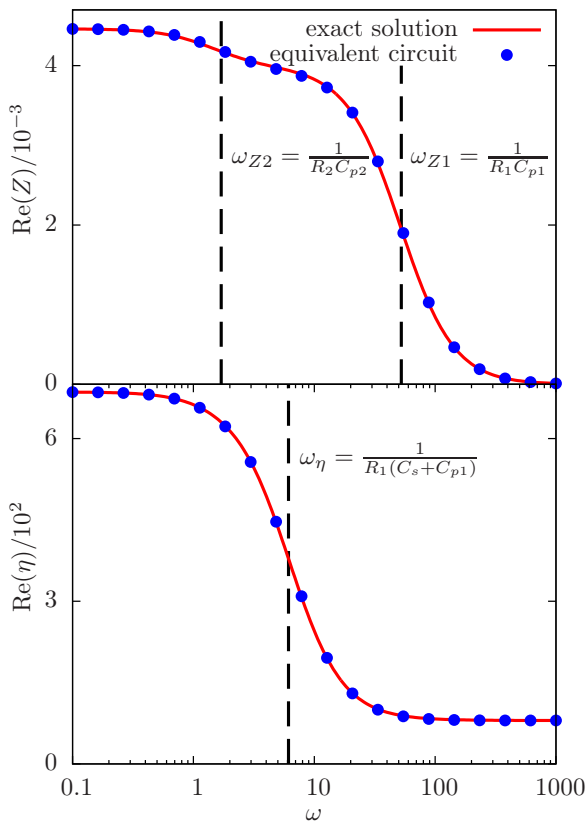}
  \caption{(Color online) Frequency dependence of the real parts of the reduced impedance $Z$ and of the relative permittivity $\eta$ for general homogeneous cells.
           There are only small deviations between the exact solution of Sec. \ref{subsubsec:The_general_homogeneous_cell} (solid red lines) and
           the equivalent circuit approximation Fig. \ref{fig:equivalent_circuit_general_homogeneous_cell} and Eq.~(\ref{eq:components_circuit_general_homogeneous_cell}) (blue dots).
           The dashed vertical lines refer to frequencies Eqs. (\ref{eq:characteristic_frequencies_general}).
           The parameters for these plots are: $L=3$, $\Phi^{st}=0$, $\bar{\varrho}=104$, $\epsilon=80$, $\Gamma_+=1$ and $\Gamma_-=2$. For the fitted parameter we found: $C_{p2}\approx 1188.31$.}
  \label{fig:spectra_homogeneous_cell_gp1_gm2_equivalent_circuit}
\end{figure}
\begin{figure}[!t]
  \includegraphics[width=0.45\textwidth]{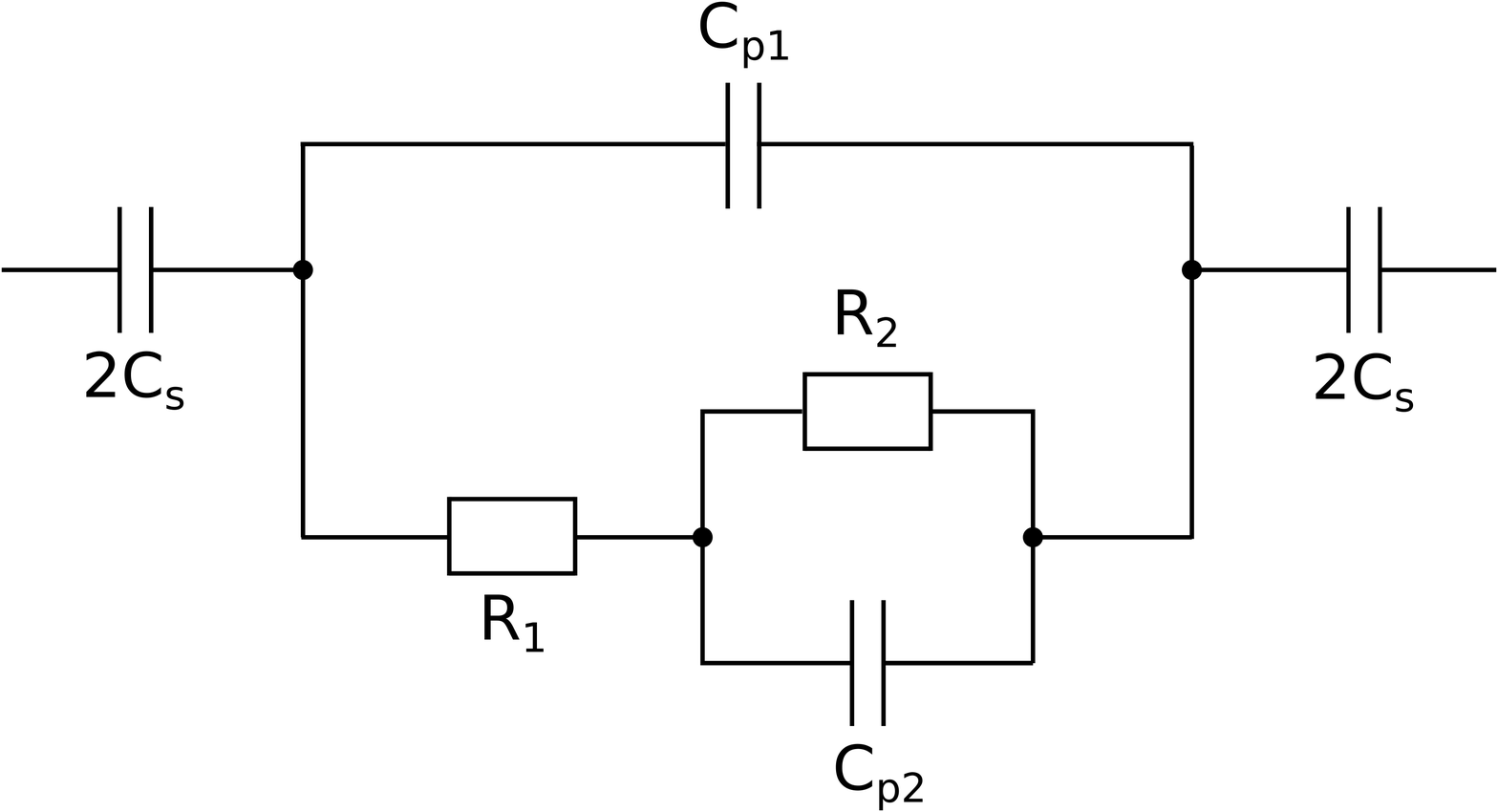}
  \caption{Equivalent circuit for general homogeneous cells consisting of Ohmic resistors $R$ and capacitors $C$. $C_s$ represents the electrodes which is why one capacitor $C_s$ has been replaced by a series circuit of two capacitors $2C_s$,
           where $2C_s$ coincides with the capacitance of a capacitor whose plates are separated by one Debye length $1/\kappa$ which is a common estimate for the range of surface effects in electrolytic solutions. The connection between the circuit elements $C_s,C_{p1},R_1$ and $R_2$ and the cell properties is given by Eqs. (\ref{eq:components_circuit_general_homogeneous_cell}). Of the five components only $C_{p2}$ has to be determined by fitting to the exact solution.}
  \label{fig:equivalent_circuit_general_homogeneous_cell}
\end{figure}

As compared to simple homogeneous cells (see Fig.~\ref{fig:spectra_homogeneous_cell_equivalent_circuit}) a pronounced new feature occurs for general homogeneous cells (see Fig.~\ref{fig:spectra_homogeneous_cell_gp1_gm2_equivalent_circuit}), in which the ion species have different mobilities $\Gamma_+\neq\Gamma_-$. At low frequencies a ``small step'' arises in $\Re(Z)$, which does not occur for the equivalent circuit of Fig.~\ref{fig:equivalent_circuit_simple_homogeneous_cell}. For that reason an extension of the latter has to be performed. The reduced impedance $Z$ of the simple homogeneous cell in the equivalent circuit approximation Eq.~(\ref{eq:Impedance_circuit_simple_homogeneous}) may be written as
\begin{align}
  Z(\omega)=\frac{b_1^{\ast}\text{i}\omega+b_0^{\ast}}{\text{i}\omega(a_1^{\ast}\text{i}\omega+a_0^{\ast})},
  \label{eq:Impedance_circuit_simple_homogeneous_polynomial}
\end{align}
where the coefficients $a_j^{\ast}$ and $b_j^{\ast}$ with $j\in\{0,1\}$ are dependent on the components $R_p,C_p$ and $C_s$. A convenient extension of the expression Eq.~(\ref{eq:Impedance_circuit_simple_homogeneous_polynomial}) might be the addition of terms quadratic in $(\text{i}\omega)$ to the complex polynomials in the numerator and denominator
\begin{align}
  Z(\omega)=\frac{b_2(\text{i}\omega)^2+b_1\text{i}\omega+b_0}{\text{i}\omega(a_2(\text{i}\omega)^2+a_1\text{i}\omega+a_0)}.
  \label{eq:Impedance_circuit_general_homogeneous_polynomial}
\end{align}
In order to derive the structure of an equivalent circuit with help of Eq.~(\ref{eq:Impedance_circuit_general_homogeneous_polynomial}) the rational function can be rewritten 
by means of an expansion in partial fractions and polynomial division
such that Ohmic resistors and capacitors may be identified based on their 
characteristic frequency dependencies. The result is an extended equivalent circuit which is shown in Fig.~\ref{fig:equivalent_circuit_general_homogeneous_cell}.
Whereas the reduced impedance of general homogeneous cells is analytically available in principle (see Sec. \ref{subsubsec:The_general_homogeneous_cell}), the
expression is too complex for a discussion similar to Sec. \ref{subsubsec:equivalent_circuit_for_simple_homogeneous_cells}.
Nevertheless it was possible to find empirical expressions for nearly all of them.
Compared to the equivalent circuit of the simple homogeneous cell (see Fig.~\ref{fig:equivalent_circuit_simple_homogeneous_cell})
the parallel circuit $R_2\parallel C_{p2}$ is the only structural change in the circuit Fig.~\ref{fig:equivalent_circuit_general_homogeneous_cell}. Analytic expressions for the remaining components, for which corresponding components exist in the circuit of the simple cell, could be found based on Eq.~(\ref{eq:components_circuit_simple_homogeneous_cell}). There, $C_s$ and $C_p$ are independent of the mobility $\Gamma$ which is the only parameter differing the simple and the general homogeneous cell. Therefore we identify the general circuit components 
$C_s$ and $C_{p1}$ with the simple circuit components $C_s$ and $C_p$, respectively.
The Ohmic resistor $R_p$ depends on the mobility. In order to obtain an expression for its corresponding component $R_1$ in the general cell
we replace the mobility $\Gamma$ by the mean value $(\Gamma_++\Gamma_-)/2$. To identify the Ohmic resistor $R_2$ we profit from the
observation that for $\omega\rightarrow0$ and $\omega\rightarrow\infty$ the reduced impedance of the general homogeneous cell
$Z(\Gamma_+,\Gamma_-)$ approaches the mean value of the reduced impedances of two simple homogeneous cells $Z(\Gamma)$ in the following way:
\begin{align}
  Z(\Gamma_+,\Gamma_-)\approx\frac{1}{2}\big(Z(\Gamma_+)+Z(\Gamma_-)\big)\quad\text{for}\quad\omega\rightarrow0,\infty
  \label{eq:Impedance_general_homogeneous_estimate}
\end{align}
For $\omega\rightarrow0$ the real part of the circuit Fig.~\ref{fig:equivalent_circuit_general_homogeneous_cell} is $\Re(Z(\omega=0))=R_1+R_2$. With Eq.~(\ref{eq:Impedance_general_homogeneous_estimate}) an estimate for $\Re(Z(\omega=0))$ is available and an expression for $R_1$ is already known. Therefore an expression for $R_2$ may be derived. 
For $L\to\infty$ the components of Fig.~\ref{fig:equivalent_circuit_general_homogeneous_cell} are estimated by
 \begin{align}
  \begin{aligned}
   C_s   &\simeq \sqrt{\frac{\epsilon \bar{\varrho}}{2\pi}}\\
   C_{p1}&\simeq \frac{\epsilon}{2\pi L}\\
   R_1   &\simeq \frac{L}{2\bar{\varrho}(\Gamma_++\Gamma_-)}\\
   R_2   &\simeq \frac{L}{8\bar{\varrho}}\frac{(\Gamma_+-\Gamma_-)^2}{\Gamma_+\Gamma_-(\Gamma_++\Gamma_-)}
  \end{aligned}
  \label{eq:components_circuit_general_homogeneous_cell}
 \end{align}
whereas the value of $C_{p2}$ has to be determined by fitting to the exact solution.

Obviously the circuit in Fig.~\ref{fig:equivalent_circuit_general_homogeneous_cell} with the elements Eq.~(\ref{eq:components_circuit_general_homogeneous_cell}) contains the special case of the simple homogeneous cell ($\Gamma_+=\Gamma_-$), where $R_2$ vanishes (see Fig.~\ref{fig:equivalent_circuit_simple_homogeneous_cell}).
However, the special case of the one component plasma, where \emph{large} differences in the ion mobilities occur, requires more sophisticated circuits.
Figure~\ref{fig:spectra_homogeneous_cell_gp1_gm2_equivalent_circuit} shows the
excellent agreement of the spectra of the empirical equivalent circuit in Fig.~\ref{fig:equivalent_circuit_general_homogeneous_cell} and Eq.~(\ref{eq:components_circuit_general_homogeneous_cell}) with those of the exact solution in Sec. \ref{subsubsec:The_general_homogeneous_cell}. The mobilities $\Gamma_+$ and $\Gamma_-$ differ by a factor of two, which is a typical ratio for monovalent ions in water \cite{Li1974}.
Figure~\ref{fig:spectra_homogeneous_cell_gp1_gm2_equivalent_circuit} exhibits
crossover phenomena at frequencies (see the dashed vertical lines) which can be estimated by
\begin{align}
  \begin{aligned}
    &\omega_{\eta}=\frac{1}{R_1(C_s+C_{p1})},\\
    &\omega_{Z1}=\frac{1}{R_1C_{p1}},\\
    &\omega_{Z2}=\frac{1}{R_2C_{p2}}.
  \end{aligned}
  \label{eq:characteristic_frequencies_general}
\end{align}
Here $\omega_{\eta}$ and $\omega_{Z1}$ correspond to the same crossover frequencies as their simple cell equivalents Eqs. (\ref{eq:characteristic_frequency_ReZ_homogeneous},\ref{eq:characteristic_frequency_Reeta_homogeneous}), whereas the crossover frequency $\omega_{Z2}$ corresponds to the frequency of the small step in $\Re(Z)$.\\
The parallel circuit $R_2\parallel C_{p2}$ can be interpreted as a correction to the simple case. In Fig.~\ref{fig:equivalent_circuit_simple_homogeneous_cell} the interior of a simple cell was described by $R_p$ and $C_p$ both of which may be explained in terms of a homogeneous field in the cell (see Eq.~(\ref{eq:components_circuit_simple_homogeneous_cell})). Such a description breaks down in the general case, where the ions with the higher mobility are able to react faster on the electric field with the result that they screen the field for the slower ions.

\subsubsection{\label{subsubsec:equivalent_circuit_for_simple_cells_with_interface}Equivalent circuit for simple cells with an interface}
\begin{figure}[!t]
  \includegraphics[width=0.45\textwidth]{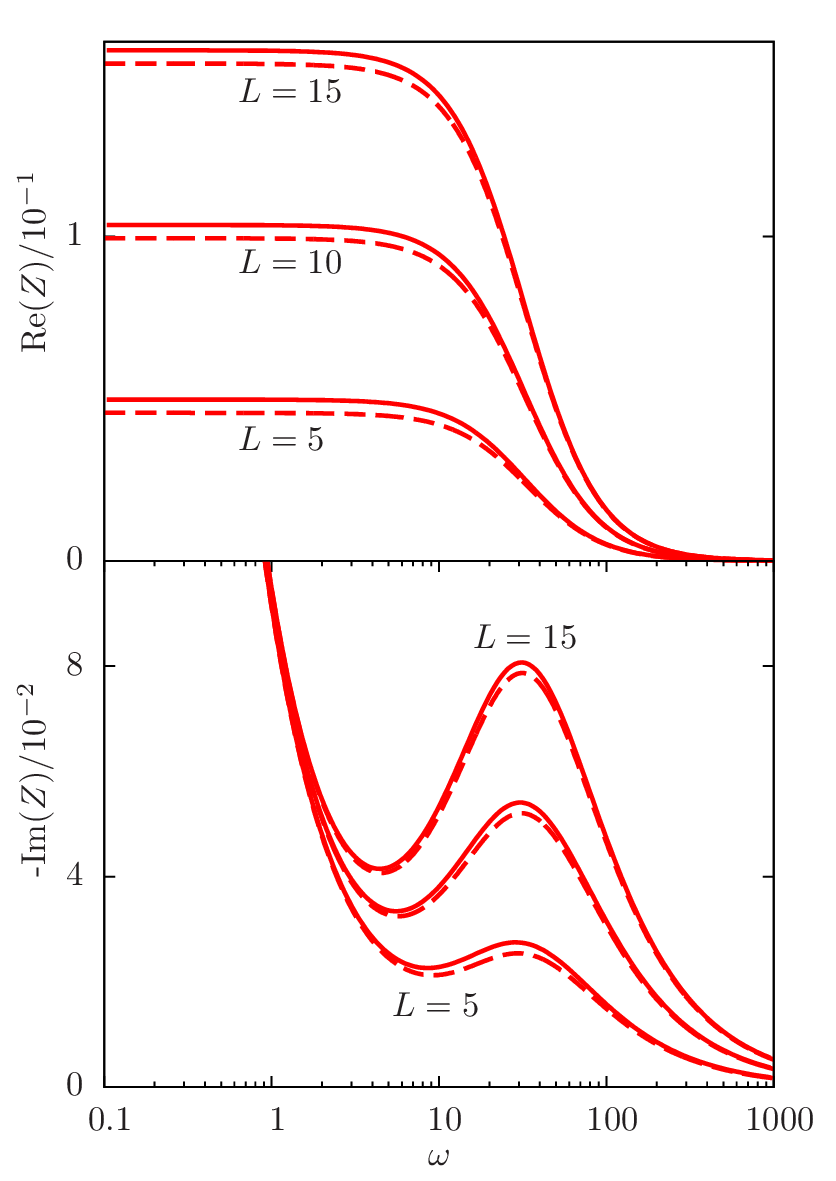}
  \caption{(Color online) Reduced impedance spectra of simple cells with an interface calculated by the method of
           Sec.~\ref{subsubsec:Cell_with_interface_equal_partitioning} (solid lines) in comparison with spectra of an equivalent circuit
           (dashed lines) similar to the one in Fig. \ref{fig:equivalent_circuit_simple_interface_cell} but 
           without the parallel subcircuit representing the interface. All elements are determined by 
           Eqs.~(\ref{eq:components_circuit_simple_homogeneous_cell}).
           As the interface contribution does not scale with the cell length $L$ its relative influence decreases with increasing cell length.
           The common parameters for these plots are: $\Phi^{st}=0$, $R=S=L/2$, $\epsilon_L=80$, $\epsilon_R=10$, $f_\pm=2$, $\bar{\varrho}=104$,
           $\Gamma_\pm^L=1$ and $\Gamma_\pm^R=0.5$.}
  \label{fig:influence_of_the_interface}
\end{figure}
\begin{figure}[!t]
  \includegraphics[width=0.45\textwidth]{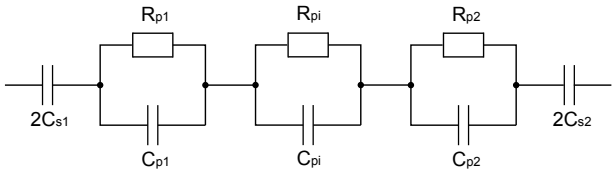}
  \caption{Equivalent circuit for simple cells with an interface consisting of Ohmic resistors $R$ and capacitors $C$. The six components of the partial cells (indices $1,2$) are given by the analytic expressions Eq.~(\ref{eq:components_circuit_simple_homogeneous_cell}). The interface is represented by the parallel circuit $R_{pi}\parallel C_{pi}$. Both components in general have to be determined by fitting to the exact solution. (See main text for further information.)}
  \label{fig:equivalent_circuit_simple_interface_cell}
\end{figure}
\begin{figure}[!t]
  \includegraphics[width=0.45\textwidth]{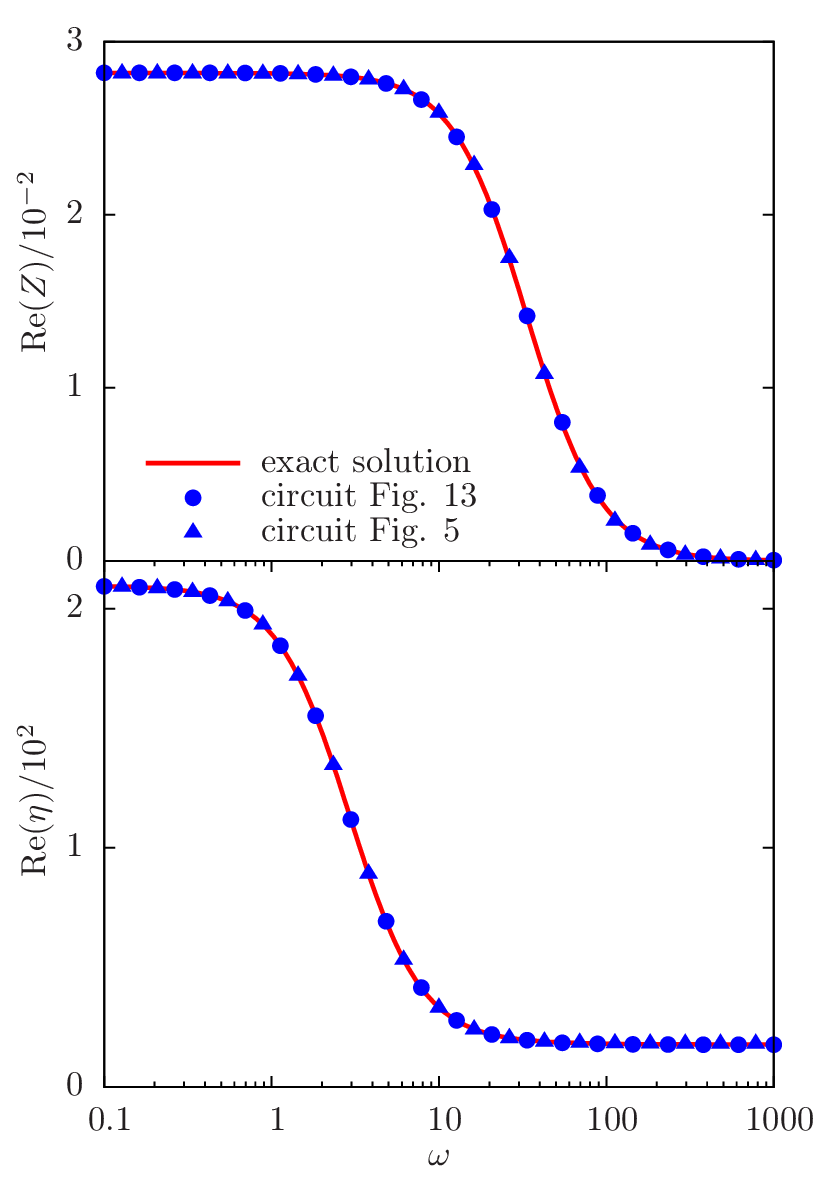}
  \caption{(Color online) Frequency dependence of the real parts of the reduced impedance $Z$ and of the relative permittivity $\eta$ for simple cells with an
           interface. The comparison between the exact solution of Sec.~\ref{subsubsec:Cell_with_interface_equal_partitioning}
           (solid red lines), the equivalent circuit approximation of Fig.~\ref{fig:equivalent_circuit_simple_interface_cell} with the two-step
           fitting approach (blue circular dots) and 
           of Fig.~\ref{fig:equivalent_circuit_simple_homogeneous_cell} (blue triangular dots) shows overall
           quantitative agreement of the spectra. The parameters for these plots are
           $L=3$, $\Phi^{st}=0$, $R=S=L/2$, $\epsilon_L=80$, $\epsilon_R=10$, $f_\pm=2$, $\bar{\varrho}=104$, $\Gamma_\pm^L=1$
           and $\Gamma_\pm^R=0.5$. The fitted interface components of
           Fig.~\ref{fig:equivalent_circuit_simple_interface_cell} are given
           by $R_{pi}\approx4.11\cdot10^{-3}$, $C_{pi}\approx10.51$, whereas
           the fitted elements of the homogeneous cell circuit
           Fig.~\ref{fig:equivalent_circuit_simple_homogeneous_cell} are 
           $C_s\approx11.12$, $R_p\approx2.82\cdot10^{-2}$ and
           $C_p\approx1.05$.
           Although the three spectra agree quantitatively, fitting the circuit Fig.~\ref{fig:equivalent_circuit_simple_homogeneous_cell}
           is not appropriate to analyze interfacial processes, since the circuit elements lump bulk and interfacial contributions
           together.
           In contrast, the proposed two-step method leads to a unique and transparent separation between bulk and interfacial 
           properties.}
  \label{fig:spectra_interface_cell_equivalent_circuit}
\end{figure}

Having discussed possible equivalent circuits for homogeneous cells in the previous 
Secs.~\ref{subsubsec:equivalent_circuit_for_general_homogeneous_cells} and \ref{subsubsec:equivalent_circuit_for_simple_homogeneous_cells}, 
a general strategy to obtain equivalent circuits and the interfacial elements for cells with a liquid-liquid interface is described here.
For the sake of clarity the discussion shall be restricted to the most simple case of an interface between two \emph{simple} homogeneous cells, 
i.e., with equal ion mobilities $\Gamma_+(z)=\Gamma_-(z)$. 
Fitting equivalent circuits to experimental data in order to extract bulk 
\emph{as well as} interfacial quantities simultaneously one has to reckon with a huge uncertainty for the latter, since their contributions to 
the impedance and the relative permittivity scale only with the interfacial area, whereas the bulk quantities scale with the cell volume.
In Fig.~\ref{fig:influence_of_the_interface} calculated impedance spectra (solid lines) are shown for different cell lengths $L$.
They are compared with the spectra of an equivalent circuit (dashed lines) in which two bulk phases Fig.~\ref{fig:equivalent_circuit_simple_homogeneous_cell} with the elements
chosen according to Eqs.~(\ref{eq:components_circuit_simple_homogeneous_cell})
and the electrode contributions ($2C_s$) are connected in series, but in which no
contributions of the liquid-liquid interface are considered (compare with Fig.~\ref{fig:equivalent_circuit_simple_interface_cell}).
Obviously, with increasing system size $L$, the impedance spectra become more and more dominated by the bulk
contributions, which scale linearly with $L$, as compared to the interfacial contributions (the difference 
between the solid and the dashed lines), which are independent of $L$.
Hence in complex equivalent circuits the precision of the bulk elements can be expected to be increased at the expense of that of the interfacial 
elements.
The relatively small contribution to an impedance spectrum of a liquid-liquid interface as compared to the bulk phases
is in line  with the earlier observation of qualitatively similar impedance spectra for simple cells with and without an interface (see
Fig.~\ref{fig:spectra_with_and_without_interface}). Consequently it is
possible to fit the equivalent circuit of a simple \emph{homogeneous} cell to the spectra of a simple cell \emph{with} interface 
(see Fig. \ref{fig:spectra_interface_cell_equivalent_circuit}), which leads to the contributions of two bulk phases and one
liquid-liquid interface lumped together in two elements $R_p$ and $C_p$. 
However, it is not possible to extract the interfacial characteristics from these lumped elements in a unique and transparent way.
As a solution of this problem we propose to perform the determination of interfacial quantities in two steps: First the equivalent circuits of the homogeneous partial 
cells, of which the bulk of the cell is composed are fixed. In the second step a series of
the previously determined bulk circuits and an additional equivalent circuit corresponding to the interface is used to fit the interfacial 
elements only.
This approach is analogous to the method of determining the interfacial tension from the total free energy by first subtracting the bulk 
contribution. Since the
bulk elements in Fig.~\ref{fig:equivalent_circuit_simple_interface_cell} are fixed during the second step, a compensation of errors as described above is not possible.
The six bulk components of the partial cells (indices $1$ and $2$) are 
chosen according to the analytic expressions Eq.~(\ref{eq:components_circuit_simple_homogeneous_cell}).
A systematic search for the smallest and quantitatively satisfying equivalent circuit representing the interface led to
the parallel circuit $R_{pi}\parallel C_{pi}$ whose components are the only ones which in general have to be
determined by fitting to the spectra of the exact solution. 

\begin{figure}[!t]
  \includegraphics[width=0.45\textwidth]{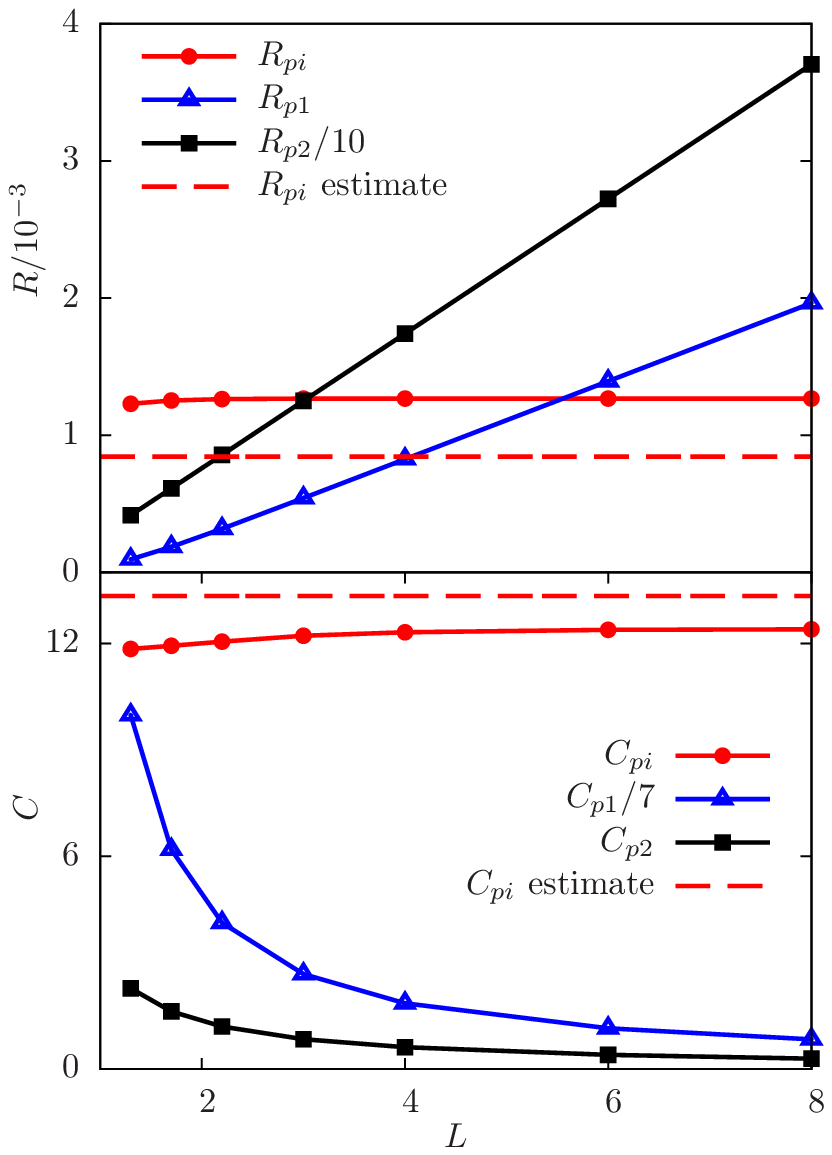}
  \caption{(Color online) Components of the equivalent circuit in Fig.~\ref{fig:equivalent_circuit_simple_interface_cell} as a function of cell length $L$. The estimates were determined by Eq.~(\ref{eq:estimation_interface_components}). Note that some of the curves shown have been weighted by constant factors. The common parameters of the plots are $\bar{\varrho}=104$, $\Phi^{st}=0$, $\epsilon_L=80$, $\epsilon_R=10$, $f_\pm=2$, $R=S\approx0.3L$ and $\Gamma^{L,R}_\pm=1$.}
 \label{fig:components_L}
\end{figure}

\begin{figure}[!t]
  \includegraphics[width=0.45\textwidth]{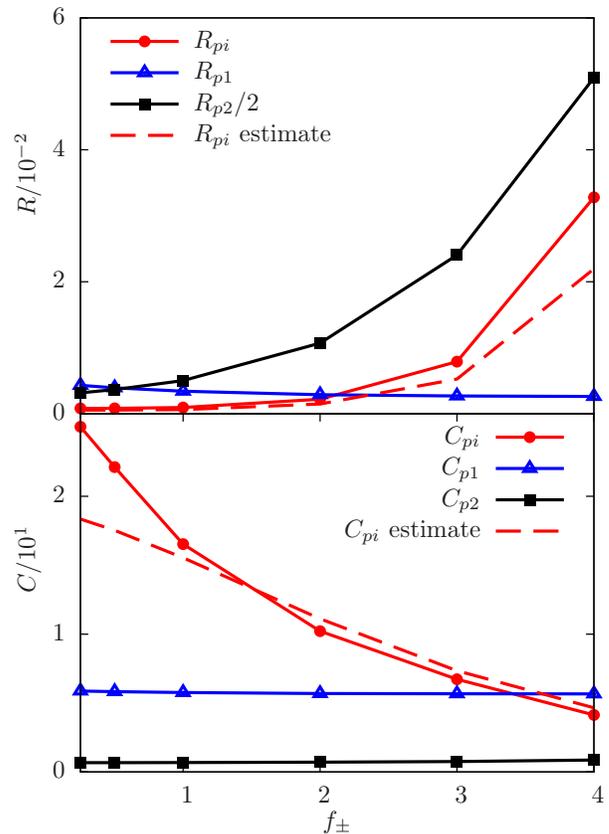}
  \caption{(Color online) Components of the equivalent circuit in Fig.~\ref{fig:equivalent_circuit_simple_interface_cell} as a function of the ion solubilities $f_\pm$. The estimates were determined by Eq.~(\ref{eq:estimation_interface_components}). Note that some of the curves shown have been weighted by constant factors. The common parameters of the plots are $L=5$, $\bar{\varrho}=104$, $\Phi^{st}=0$, $\epsilon_L=80$, $\epsilon_R=10$, $R=S=L/2$ and $\Gamma^{L,R}_\pm=1$.}
 \label{fig:components_fpm}
\end{figure}

\begin{figure}[!t]
  \includegraphics[width=0.45\textwidth]{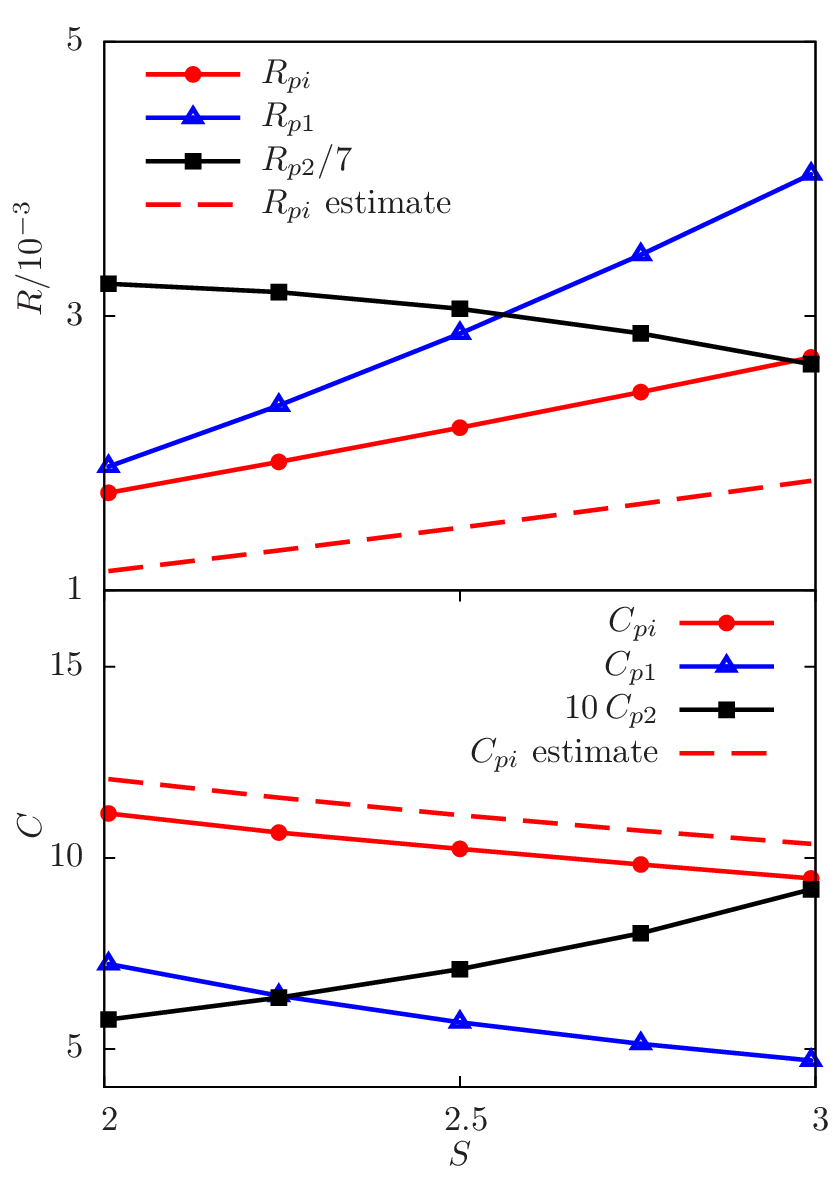}
  \caption{(Color online) Components of the equivalent circuit in Fig.~\ref{fig:equivalent_circuit_simple_interface_cell} as a function of the common position $S$ of both the $\epsilon$- and the $V_\pm$-interface (see Eqs. (\ref{eq:solvent_dielectricity},\ref{eq:external_potential})). The estimates were determined by Eq.~(\ref{eq:estimation_interface_components}). Note that some of the curves shown have been weighted by constant factors. The common parameters of the plots are $L=5$, $\bar{\varrho}=104$, $\Phi^{st}=0$, $\epsilon_L=80$, $\epsilon_R=10$, $f_\pm=2$ and $\Gamma^{L,R}_\pm=1$.}
 \label{fig:components_R}
\end{figure}

\begin{figure}[!t]
  \includegraphics[width=0.45\textwidth]{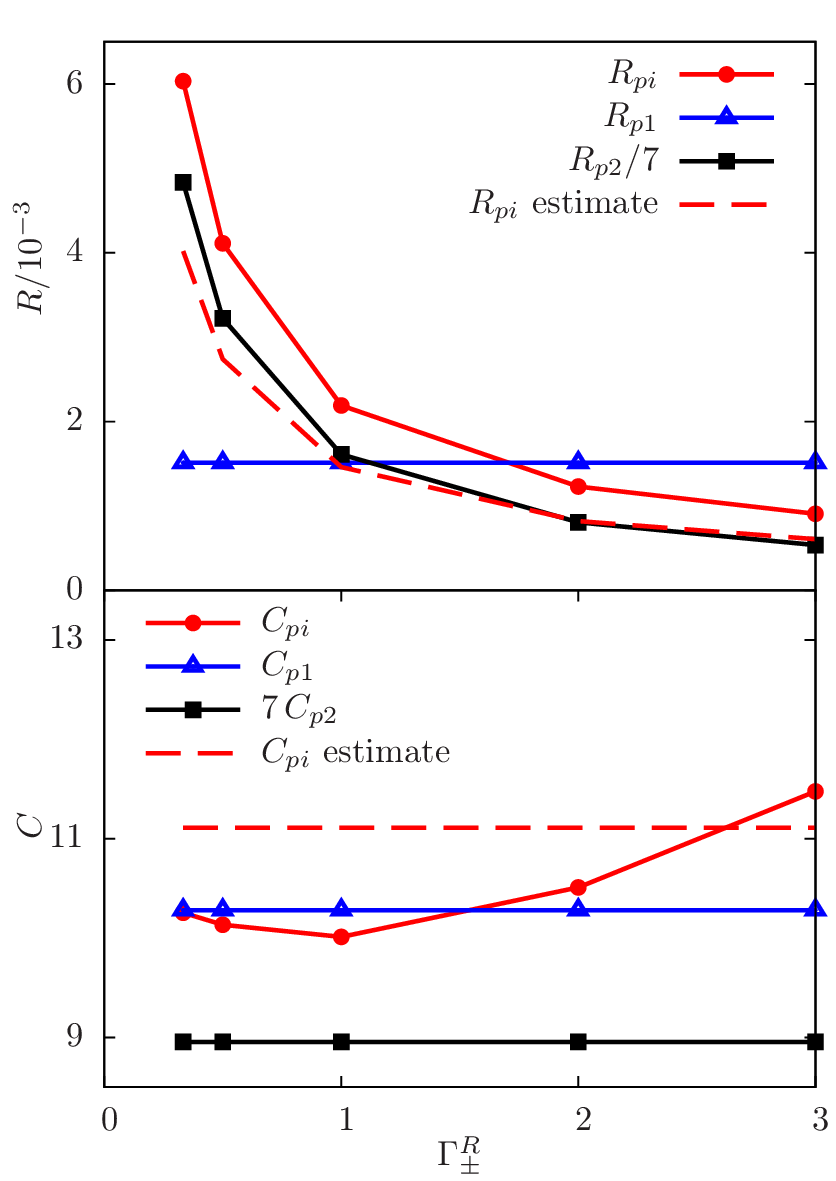}
  \caption{(Color online) Components of the equivalent circuit in Fig.~\ref{fig:equivalent_circuit_simple_interface_cell} as a function of the mobility $\Gamma^R_\pm$ in the right partial cell. The estimates were determined by Eq.~(\ref{eq:estimation_interface_components}). Note that some of the curves shown have been weighted by constant factors. The common parameters of the plots are $L=3$, $\bar{\varrho}=104$, $\Phi^{st}=0$, $\epsilon_L=80$, $\epsilon_R=10$, $f_\pm=2$, $R=S\approx L/2$ and $\Gamma^{L}_\pm=1$.}
 \label{fig:components_gr}
\end{figure}

The interfacial elements $R_{pi}$ and $C_{pi}$ displayed as red lines with circles in Figs. \ref{fig:components_L}-\ref{fig:components_gr} have been obtained by fitting the spectra of the equivalent circuit Fig. \ref{fig:equivalent_circuit_simple_interface_cell} to exact spectra obtained by the method in Sec. \ref{subsubsec:Cell_with_interface_equal_partitioning}.\\
The interface elements $R_{pi}$ and $C_{pi}$ are the only ones for which in general no analytic expressions are available. However, estimates are obtained by forming series circuits out of cell elements for simple homogeneous cells Eq.~(\ref{eq:components_circuit_simple_homogeneous_cell}) with lengths $1/\kappa_{L,R}$:
\begin{align}
 \begin{aligned}
  R_{pi}&\approx\left(4\bar{\varrho}_L\kappa_L\Gamma^{L}_\pm\right)^{-1}+\left(4\bar{\varrho}_R\kappa_R\Gamma^{R}_\pm\right)^{-1}\\
  C_{pi}&\approx\left(\sqrt{\frac{\pi}{2\epsilon_L \bar{\varrho}_L}}+\sqrt{\frac{\pi}{2\epsilon_R \bar{\varrho}_R}}\right)^{-1}
 \end{aligned}
 \label{eq:estimation_interface_components}
\end{align}
These estimates are displayed as dashed red lines in Figs. \ref{fig:components_L}-\ref{fig:components_gr}.
\paragraph{Dependency on the cell length.}
In Fig.~\ref{fig:components_L} only the components of the simple homogeneous partial cells exhibit dependencies on the cell length $L$,
whereas the
interface elements are virtually constant. This observation is not unexpected: features of the interface should be independent of the cell length.
Only for small $L$ we notice deviations from the constant behavior, 
which can be understood by the mutual influence of the interface with the electrodes at close separations.
The estimates, of course, do not take into account these overlap effects which is why they are independent of $L$. When ignoring the behavior for small $L$ they differ from the red lines with circles only by a constant factor close to unity.

\paragraph{Dependency on the ion solubilities.}
Figure~\ref{fig:components_fpm} displays the resulting elements of the equivalent circuit of Fig.~\ref{fig:equivalent_circuit_simple_interface_cell} for different values of the ion solubilities $f_+=f_-$. As the ion solubilities affect the ionic strengths in Eq.~(\ref{eq:ionic_strengths}) the Ohmic resistances $R_{p1,2}$ of the partial cells show a dependence on $f_\pm$ in contrast to the capacitors $C_{p1,2}$. The estimates of the interface elements are associated with the Debye lengths which is why both $R_{pi}$ and $C_{pi}$ are dependent on the solubilities. In the case of $R_{pi}$ the estimate mainly differs from the red line with circles by a constant factor whereas in the case of $C_{pi}$ there are qualitative differences for small $f_\pm$. The Ohmic resistance $R_{pi}$ attains relatively small values in the considered range so fitting the value $C_{pi}$ is inaccurate.

\paragraph{Dependency on the position of the interface.}
Figure~\ref{fig:components_R} depicts the circuit elements
upon varying the common position $S$ of both the $\epsilon$- and the $V_\pm$-interface (see Eqs. (\ref{eq:solvent_dielectricity},\ref{eq:external_potential})).
As the average ion density $\bar{\varrho}$ is held constant in all configurations the ionic strengths in the partial cells are functions of $S$, see Eq.~(\ref{eq:ionic_strengths}). Moreover, the elements of the partial cells are functions of the lengths of the partial cells $S$ and $L-S$, respectively. One would expect the interface elements to be independent of the interface position. But of course the changing ionic strengths affect the estimates Eq.~(\ref{eq:estimation_interface_components}). Because they mainly differ by constant factors from the red lines with circles the cause of the $S$-dependence
can be ascribed to the changing ionic strengths. This in turn confirms the expectation that the interface properties are indeed independent of the interface position. 
However, this is true only for sufficiently large distances between the interface and the electrodes, for which the mutual influence is negligible.

\paragraph{Dependency on the mobility.}
Now the mobility of both ion species in the left partial cell is assumed to be 
fixed to $\Gamma^L_\pm=1$,
whereas the mobilities $\Gamma^R_\pm$ in the right partial cell are varied. In Fig.~\ref{fig:components_gr} the resulting circuit elements are summarized. In the case of the Ohmic resistance the estimate merely differs from the red line with circles by a constant factor. However, the estimate for the interfacial capacitor deviates qualitatively. The prediction, following Eq.~(\ref{eq:estimation_interface_components}), inherits independence of the mobilities. This is why the dashed line in Fig.~\ref{fig:components_gr} is constant. By contrast the red line with circles shows a dependence with a minimum at $\Gamma^R_\pm=1$. The observation might be explained as follows: unequal mobilities in the left and the right partial cells lead to 
jamming
of the ions at the interface because the ions in the faster phase have to ``wait'' for ions in the slower phase to create space. This picture leads to the conclusion that bigger differences in the mobilities 
lead to an accumulation of more charge at the interface. In the equivalent circuit the interfacial charge is represented by the capacitor $C_{pi}$: A
larger
capacitance corresponds to a larger charge accumulation.

\section{\label{sec:Conclusion}Conclusion and summary}
In the previous Sec.~\ref{subsubsec:equivalent_circuit_for_simple_cells_with_interface} we proposed
a novel approach to infer properties of interfaces between two ion conducting liquids 
from impedance spectra.
For reasons of simplicity and clarity we studied idealized electrochemical cells (see 
Fig.~\ref{fig:sketch}) with blocking electrodes whose ion dynamics were described within a dynamic
density functional theory (see Eq.~(\ref{eq:free_energy_functional})) which is equivalent to the
Nernst-Planck-Poisson theory (see Eqs.~(\ref{eq:Poisson_equation},\ref{eq:Nernst_Planck_equation})).
The solutions, i.e., the ion densities $\varrho_\pm(z,t)$ and the electric potential $\phi(z,t)$,
were used to determine the reduced impedance $Z(\omega)$ (see Eq.~(\ref{eq:impedance})) and the
relative permittivity $\eta(\omega)$ (see Eq.~(\ref{eq:dielectricity})).
A concise way to represent impedances is in terms of equivalent circuits, whose elements are
commonly assigned to certain microscopic processes in the bulk as well as at the interface.
However, instead of fitting a complex equivalent circuit in one step to the calculated relative
impedance $Z(\omega)$ (see, e.g., Figs.~\ref{fig:spectra_with_and_without_interface} and 
\ref{fig:spectra_interface_cell_equivalent_circuit}),
we proposed a two-step fitting procedure,
where the bulk properties of the two partial cells have to be determined first and fixed afterwards
while fitting the interfacial properties.
We demonstrated this approach for the case of simple cells, for which the local mobilities of the
cations and anions are equal.
For this case, the bulk of the partial cells can be quantitatively represented by the equivalent
circuit displayed in Fig.~\ref{fig:equivalent_circuit_simple_homogeneous_cell} (see 
Fig.~\ref{fig:spectra_homogeneous_cell_equivalent_circuit}).
Fixing the bulk elements of the equivalent circuit 
Fig.~\ref{fig:equivalent_circuit_simple_interface_cell}, the interfacial elements are
then fitted to the calculated relative impedance $Z(\omega)$ (see 
Fig.~\ref{fig:spectra_interface_cell_equivalent_circuit}).
The bulk and interfacial elements obtained by this method exhibit intuitive dependencies on the
cell size (Fig.~\ref{fig:components_L}), solubilities (Fig.~\ref{fig:components_fpm}), interfacial
position (Fig.~\ref{fig:components_R}), and mobilities (Fig.~\ref{fig:components_gr}).
This is not necessarily the case when applying a one-step fitting procedure, in which 
an increased accuracy of the bulk elements at the expense of a decreased
accuracy of the interfacial elements may occur.
However, the main goal of the proposed two-step method is to uniquely assign equivalent
circuit elements to either bulk or interfacial processes, while the more traditional one-step 
approach does not guarantee this.
The two-step procedure we are proposing here turns out to be numerically robust as well as
intuitively appealing such that it can be expected to be useful for the interpretation of
impedance spectra. 

%\begin{acknowledgments}
%We thank ...
%\end{acknowledgments}

%-------------------------------------------------------------------------------

% \appendix

%-------------------------------------------------------------------------------

%-------------------------------------------------------------------------------

\end{document}